\documentclass[draft]{agujournal2019}
\usepackage{url} 
\usepackage{lineno}
\usepackage[inline]{trackchanges} 
\usepackage{soul}
\usepackage{amsmath}
\draftfalse

\journalname{Geophysical Research Letters}

\begin{document}

%
%

\title{Distribution of interseismic coupling along the North and East Anatolian Faults inferred from InSAR and GPS data}

%
%




\authors{Quentin Bletery$^1$, Olivier Cavali\'e$^1$, Jean-Mathieu Nocquet$^{1,2}$ and  Th\'ea Ragon$^3$}

\affiliation{1}{Universit\'e C\^ote d'Azur, IRD, CNRS, Observatoire de la C\^ote d'Azur, G\'eoazur, France}
\affiliation{2}{Institut de Physique du Globe de Paris, Universit\'e de Paris, CNRS, France}
\affiliation{3}{Seismological Laboratory, California Institute of Technology, USA}






\correspondingauthor{Quentin Bletery}{bletery@geoazur.unice.fr}




\begin{keypoints}
\item Distribution of interseismic coupling on the North and East Anatolian Faults 
\item Quantification of uncertainties on coupling and Euler poles in a Bayesian inversion framework
\item The 2020 $\rm{M_w}$ 6.8 Elaz{\i}\u{g} earthquake released 221.5 years ($\pm$ 26) of accumulated moment
\end{keypoints}

%
%

%
%


\begin{abstract}
The North Anatolian Fault (NAF) has produced numerous major earthquakes. After decades of quiescence, the $\rm{M_w}$ 6.8 Elaz{\i}\u{g} earthquake (January 24, 2020) has recently reminded us that the East Anatolian Fault (EAF) is also capable of producing significant earthquakes. To better estimate the seismic hazard associated with these two faults, we jointly invert Interferometric Synthetic Aperture Radar (InSAR) and GPS data to image the spatial distribution of interseismic coupling along the eastern part of both the North and East Anatolian Faults. We perform the inversion in a Bayesian framework, enabling to estimate uncertainties on both long-term relative plate motion and coupling. We find that coupling is high and deep (0-20 km) on the NAF and heterogeneous and superficial (0-5 km) on the EAF. Our model predicts that the Elaz{\i}\u{g} earthquake released between 200 and 250 years of accumulated moment, suggesting a bi-centennial recurrence time.
\end{abstract}

\section*{Plain Language Summary}
Earthquakes are thought to occur on coupled fault portions, which are ``locked'' during the time separating two earthquakes while tectonic plates are steadily moving. The spatial distribution of coupling has been imaged along numerous large faults in the world, but despite its considerable associated seismic hazard, not on the North Anatolian Fault (NAF). The recent $\rm{M_w}$ 6.8 Elaz{\i}\u{g} earthquake (January 24, 2020) has reminded us that the East Anatolian Fault (EAF) is also capable of producing large earthquakes. To better assess the seismic hazard associated with both the NAF and the EAF, we image the distribution of interseismic coupling along these faults. We find that the NAF is strongly coupled along most of the studied section. On the opposite, coupling is shallow and heterogeneous along the EAF. The initiation of the Elaz{\i}\u{g} earthquake coincides with a strongly locked but narrow (5 x 14 km) and superficial patch. The rest of the rupture extends over moderately coupled fault portions. We estimate that it took between 200 and 250 years to accumulate the moment released by the Elaz{\i}\u{g} event. Several fault segments along the EAF present similar coupling distributions, suggesting that, provided enough time, they could host earthquakes of similar magnitude.

%
%

%


%
%
%
%

\section{Introduction}

\begin{figure}
\hspace{-2cm}\noindent\includegraphics[width=18cm]{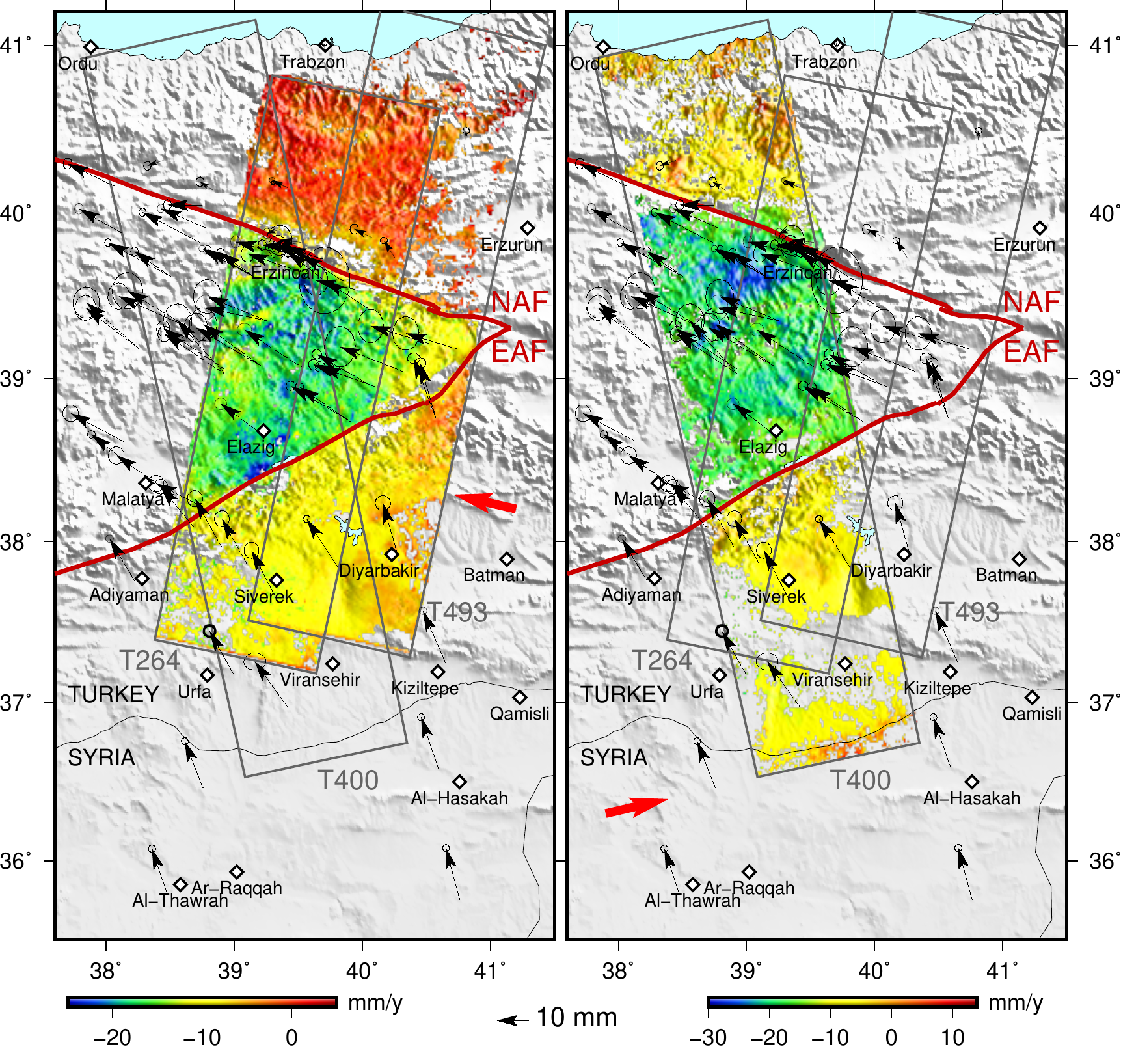}   
\caption{The NAF-EAF system (red lines) and available observations of surface deformation. Color maps show InSAR horizontal velocities (in a Eurasia‐fixed reference frame) in the satellite line of sight (LOS) direction (thick red arrows), $\sim$ $103^{\circ}$ N for descending tracks T264 and T493 (left), $\sim$ $77^{\circ}$ N for ascending track T400 (right) \cite{cavalie2014block}. Black arrows show GPS measurements and their 95\% ellipses of uncertainty \cite{nocquet2012present, ozener2010kinematics, tatar2012crustal}. White diamonds indicate large ($>100,000$ people) cities.}
\label{intro}
\end{figure}

Earthquakes are thought to rupture fault portions that have previously accumulated a deficit of slip over tens to thousands of years \cite <e.g., >[]{avouac2015geodetic}. Quantifying the spatial distribution of interseismic coupling -- i.e. the percentage of slip deficit with respect to the long-term drift of tectonic plates -- along large faults is therefore crucial to anticipate earthquakes and better assess seismic hazard \cite <e.g., >[]{kaneko2010towards}. The emergence of space geodetic techniques has allowed to infer interseismic coupling along a number of large faults during long quiescent periods of time separating one large earthquake to the next \cite <e.g., >[]{burgmann2005interseismic, moreno20102010, loveless2011spatial, protti2014nicoya, jolivet2015aseismic, metois2016interseismic, nocquet2017supercycle}. Though interseismic coupling models have been proposed to estimate the locking depth of the North and East Anatolian Faults \cite <e.g., >[]{tatar2012crustal, mahmoud2013kinematic, cavalie2014block, aktug2013seismicity, aktug2016slip}, none have quantified the lateral variations of coupling along these faults, which has limited the possibilities to study the spatial relationship between coupling and large earthquakes. The density of InSAR observations \cite{cavalie2014block} combined with sparser GPS measurements allows to infer these lateral variations of coupling on the eastern part of the NAF-EAF system (Fig. \ref{intro}).

The eastern part of the North Anatolian Fault (NAF) is known to produce large earthquakes \cite <e.g., >[]{ambraseys1971value, ambraseys1989temporary, barka1996slip} and thought to be coupled from 0 to 15 km depth \cite{reilinger2006gps, cavalie2014block}. On the other hand, simple back slip models showed that the East Anatolian Fault (EAF) is weakly coupled and only in the first kilometers of the upper crust, from 0 to 5 km \cite{cavalie2014block}. This observation was in good agreement with the relative scarcity of large earthquakes recorded during the twentieth century \cite{burton1984seismic, jackson1988relationship}. For those reasons, the January 24 2020 $\rm{M_w}$ 6.8 Elaz{\i}\u{g} earthquake came as a surprise, on a segment that does not exhibit signs of past rupture \cite{duman2013east} and in an area where the last earthquake of comparable magnitude ($\rm{M_S}$ 6.8) occurred in 1905 \cite{nalbant2002stress}. To understand this unexpected event, and more generally the seismicity in the region, we infer here the spatial distribution of interseismic coupling along the eastern part of the NAF-EAF system using InSAR \cite{cavalie2014block} and GPS measurements \cite{nocquet2012present, ozener2010kinematics, tatar2012crustal} of the interseismic surface deformation.

Inferring spatially variable interseismic coupling along faults from geodetic observations -- such as InSAR and GPS -- of the Earth surface deformation requires solving an inverse problem which usually does not admit a unique solution \cite{tarantola1982inverse, nocquet2018stochastic}. Most inversion techniques deal with this non-uniqueness by finding the solution that best fits the observations in a least square sense, together with some roughness and/or damping penalty function. As a result, typical published coupling (or slip) models are the smoothest best-fitting solutions among an infinity of possible models. We adopt here a Bayesian approach, which does not invert for a specific ``ambiguously-defined best solution'' but explores the entire solution space, sampled with respect to the likelihood of each model. This approach -- originally developed to invert for co-seismic slip models \cite{minson2013bayesian} -- also enables to reliably estimate uncertainties on coupling distributions \cite{jolivet2015aseismic, jolivet2020interseismic}.

\section{Data}

Our dataset is composed of InSAR and GPS measurements in eastern Anatolia, all calculated in a stable Eurasia reference frame (Fig. \ref{intro}). Our InSAR dataset is composed of two descending and one ascending tracks -- all crossing both the North and East Anatolian faults near their junction in eastern Turkey -- processed by \citeA{cavalie2014block}. Our GPS dataset is composed of the horizontal components of 72 GPS stations located in the area \cite{nocquet2012present, ozener2010kinematics, tatar2012crustal}.

InSAR data were derived from multiframe Envisat synthetic aperture radar images provided by the European Space Agency. Each track includes between 16 and 19 SAR images acquired between 2003 and 2010. Interferograms were generated using the New‐Small BAseline Subset (NSBAS) processing chain \cite{doin2011presentation}. They  were corrected for a ramp mostly due to a drift in the local oscillator on-board the Envisat satellite \cite{marinkovic2013consequences}. To avoid removing tectonic signals related to the motion of the Anatolian and Arabian plates, the ramps were estimated only on their Eurasian part that is considered as stable and orthogonal to the flight direction. All calculations were made considering stable Eurasia as a reference by setting the mean displacement of this area to zero, in the least squares sense. Surface displacement rates from the interferograms were derived using a small baseline time series approach, which maximizes coherence and the number of pixels to use in the analysis. A smoothing operator was applied to limit phase variations due to turbulent atmospheric delays. Finally, the linear component of the time series was extracted for each pixel in order to obtain the steady ground velocities. For a more detailed description of the InSAR processing, we refer the reader to the original study of \citeA{cavalie2014block}.

Additionally, we compiled GPS data located between longitudes 38$^{\circ}$E and 41$^{\circ}$E and latitudes 35$^{\circ}$N and 43$^{\circ}$N from 3 independent studies. Velocity for 19 points were published by \citeA{tatar2012crustal} derived from 3 surveys performed between 2006 and 2008. Another set of 19 points were published by \citeA{ozener2010kinematics} from 3 campaigns with 12-months interval. The remaining 34 points were originally published by \citeA{reilinger2006gps} and \citeA{reilinger2011nubia} but re-calculated in the continental-scale combination solution described in \citeA{nocquet2012present}. The 3 data sets are expressed in a Eurasia-fixed reference frame. The lack of enough common sites shared among the 3 solutions prevents to properly combine them, but the few common sites and analysis of models residuals does not show any systematic pattern, suggesting that the three velocity fields are consistent within their uncertainties.

\section{Bayesian inversion of rotation poles and interseismic slip deficit rate along two faults from InSAR and GPS data}

We invert the aforementioned InSAR and GPS measurements of the eastern Anatolia surface deformation to infer the distribution of interseismic slip deficit rate along the North-East Anatolian fault system using a Bayesian sampling approach implemented in the AlTar1 package, originally developed by \citeA{minson2013bayesian} under the name of CATMIP. AlTar associates Markov chain Monte Carlo methods with a tempering process to explore the solution space, each step of the tempering being followed by a resampling to select only the most probable models. The probability density function (pdf) $p(\mathbf{m}|\mathbf{d})$ of a large number of likely models $\mathbf{m}$ given our data $\mathbf{d}$ is evaluated based on the ability of a model $\mathbf{m}$ to predict the data $\mathbf{d}$ \cite{minson2013bayesian}:
\begin{linenomath*}
\begin{equation}
p(\mathbf{m}|\mathbf{d}) \propto p(\mathbf{m}) \exp [-\frac{1}{2} (\mathbf{d}-\mathbf{G}\mathbf{m})^T \mathbf{C}_{\chi}^{-1} (\mathbf{d}-\mathbf{G}\mathbf{m})],
\end{equation}
\end{linenomath*}
where $\mathbf{G}$ is the matrix of the Green's functions and $\mathbf{C}_{\chi}$ is the misfit covariance matrix. Vector $\mathbf{d}$ is composed of 144 GPS measurements (72 $\times$ 2 components) and a subset of InSAR pixels on the 3 tracks down-sampled using the Quadtree algorithm \cite{jonsson2002fault}. 

Because the inferred distribution of coupling is presumably highly sensitive to the (usually) pre-determined tectonic block motion, especially in a case involving 3 plates, we do not impose pre-calculated plate rotations but invert for them simultaneously with the interseismic slip deficit rate -- similarly to the approach proposed by \citeA{meade2009block} but adapted to a Bayesian framework. We discretize the eastern part of the North and East Anatolian faults into 110 subfaults of depth-dependent sizes (Table S1, S2) and invert for the model vector
\begin{linenomath*}
\begin{equation}
\mathbf{m} = 
\begin{pmatrix}
\mathbf{w}^1 \\
\mathbf{w}^2 \\
\mathbf{S} 
\end{pmatrix},
\end{equation}
\end{linenomath*}
where $\mathbf{w}$ is the plate rotation vector expressed in Cartesian geocentric coordinates with unit of rad/y, $^1$ stands for Anatolia with respect to Eurasia, $^2$ for Arabia with respect to Eurasia, and $\mathbf{S}$ is the back-slip on each subfault. Accordingly, we build $\mathbf{G}$ so that
\begin{linenomath*}
\begin{equation}
\mathbf{G} = 
\begin{pmatrix}
\mathbf{A}, & -\mathbf{G}_S
\end{pmatrix},
\end{equation}
\end{linenomath*}
where $\mathbf{A}$ is the matrix relating the plate rotation vectors to the horizontal velocities (see Appendix A) and $\mathbf{G}_S$ is the classical matrix of the Green's functions computed using the analytical solution of a shear finite fault embedded in an elastic half space \cite{mansinha1971displacement,okada1985surface}.

$\mathbf{C}_{\chi}$ is the misfit covariance matrix, which translates data and epistemic uncertainties into uncertainties on the inverted model $\mathbf{m}$ \cite{duputel2014accounting, bletery2016bayesian, ragon2018accounting, ragon2019accounting, ragon2019joint}. Here, we only account for data uncertainties. For GPS records, we fill $\mathbf{C}_{\chi}$ with the (squared) standard deviations and covariances between the east and north components of a given station provided in the GPS solutions. For InSAR pixels, we first remove the tectonic signal from the unsampled interferograms using a preliminary model and calculate the covariance across the pixels of the residual interferograms as a function of their distances. We fit an exponential function (Fig. S1) to the obtained cloud of points and express the covariance $C_{i,j}$ between 2 pixels as a function of their distance $D_{i,j}$
\begin{linenomath*}
\begin{equation}
C_{i,j} = a^2 \exp ( \frac{ - D_{i,j}}{b} ),
\label{cov}
\end{equation}
\end{linenomath*}
by applying a regression to the parameters $a$ and $b$ independently on the 3 tracks \cite{sudhaus2009improved, jolivet2012shallow, jolivet2015aseismic}. We then use equation \ref{cov} to evaluate the covariance on the sub-sampled interferograms.

$p(\mathbf{m})$ is the pdf describing the prior information assumed on the different model parameters. We choose the less informative distributions for back-slip parameters $\mathbf{S}$, i.e. uniform distributions between 0 and the a priori long-term interplate velocities: 19.5 mm/y for the North Anatolian and 13 mm/y for the East Anatolian fault \cite{cavalie2014block}. For the plate rotation vectors, we use the Euler poles and their associated uncertainty from \cite{le2010miocene} to derive a prior pdf. Plate rotation vectors (in Cartesian geocentric coordinates) $\mathbf{w}
^p$ are related to Euler pole parameters through
\begin{linenomath*}
\begin{equation}
\mathbf{w}^p  = \Omega^p 
\begin{pmatrix}
\cos{\phi^p} \cos{\lambda^p} \\
\cos{\phi^p} \sin{\lambda^p} \\
\sin{\phi^p}
\end{pmatrix},
\end{equation}
\end{linenomath*}
where $\lambda^p$ and $\phi^p$ are the longitude and latitude of the Euler pole of a plate $p$ and $\Omega^p$ is its angular velocity \cite{bowring1985accuracy}. Note that this change of coordinate system makes the problem linear \cite <e.g., >[]{nocquet2001intraplate,maurer2014fault,meade2009block}. We draw 100,000 sets of parameters ($\lambda^1$, $\phi^1$, $\Omega^1$, $\lambda^2$, $\phi^2$, $\Omega^2$) from normal distributions defined by means and standard deviations taken from previously published solutions \cite[summarized in Table \ref{rotation}]{le2010miocene}. For each drawn set of parameters, we calculate the corresponding $\mathbf{w}^1$ and $\mathbf{w}^2$. We obtain Gaussian-like distributions for each component of $\mathbf{w}^1$ (Fig. S2) and $\mathbf{w}^2$ (Fig. S3). We extract the mean and standard deviation of these distributions and use them to define normal prior pdfs on $\mathrm{w}_{x,y,z}^{1,2}$ in AlTar.

\begin{table}
\caption{{\it A priori} \cite{le2010miocene} and {\it a posteriori} Euler pole coordinates and angular velocities with respect to Eurasia. {\it A posteriori} parameters are the mean and 2-$\sigma$ standard deviation (95\% confidence) of the posterior pdfs (Fig. S6).}
\centering
\begin{tabular}{l | l c c c}
\hline
~ & Plate & Longitude ($^\circ$ E) & Latitude ($^\circ$ N) & Angular velocity ($^\circ$/My) \\
\hline
{\it A priori}  & Anatolia & 31.96 $\pm$ 0.10 & 32.02 $\pm$ 0.10 & 1.307 $\pm$ 0.083 \\
~ & Arabia & 15.21 $\pm$ 0.10 & 28.31 $\pm$ 0.10 & 0.396 $\pm$ 0.010 \\
\hline
{\it A posteriori}  & Anatolia & 34.22 $\pm$ 0.35 & 30.96 $\pm$ 0.60 & 1.087 $\pm$ 0.078 \\
~ & Arabia & 16.13 $\pm$ 0.52 & 27.08 $\pm$ 0.37 & 0.386 $\pm$ 0.008 \\
\hline
\end{tabular}
\label{rotation}
\end{table}

\section{Results}

We obtain a posterior marginal pdf for every inverted parameter in $\mathbf{m}$, 110 fault slip parameters and 6 parameters describing the plate rotation vectors ($\mathrm{w}_{x,y,z}^{1,2}$). The posterior pdfs on $\mathbf{w}^1$ and $\mathbf{w}^2$ parameters (Fig. S4) appear uncorrelated (coefficients of correlation $<$ 0.013) with each other and -- to a lesser extent -- with fault slip parameters (coefficients of correlation $<$ 0.13) (Fig. S5). Moderate anticorrelations are noticeable between fault slip parameters of patches located one beneath another (i.e. at the same location but different depth) (Fig. S5.a). 

We convert the inverted pdfs on the rotation vectors ($\mathbf{w}^1$, $\mathbf{w}^2$) (Fig. S4) into pdfs on the Euler pole coordinates and angular velocities (Fig. S6). The means and 2-$\sigma$ standard deviations of the inverted pdfs are summarized in Table \ref{rotation}. They are close to the previously published values we used as a prior \cite{le2010miocene} but not equal (Fig. S7). A possible explanation for this small discrepancy is that the plates are not strictly rigid \cite{le2010miocene, nocquet2012present, aktug2013seismicity, england2016constraints} and thus the rotations we invert from data in eastern Anatolia are slightly different from those obtained from data sampling a larger area of the plate. Fig. S8.a shows the velocities corrected from plate motion using the Euler poles from \citeA{le2010miocene}. It clearly shows a pattern of a residual rotation and unlikely large (5 mm/y) fault normal relative motion across both faults. Using our poles, residuals velocities appear to be consistent with the interseismic pattern (back-slip) expected for 
strike-slip faults (Fig. S8.b). Our goal here is to infer the coupling distribution, and for that aim a refined estimate of the rotation parameters close to the fault is preferable to a plate-average solution, but one should be careful in using values in Table \ref{rotation} for other purposes.

\begin{figure}
\vspace{-.5cm}\hspace{-1.5cm}\noindent\includegraphics[width=17cm]{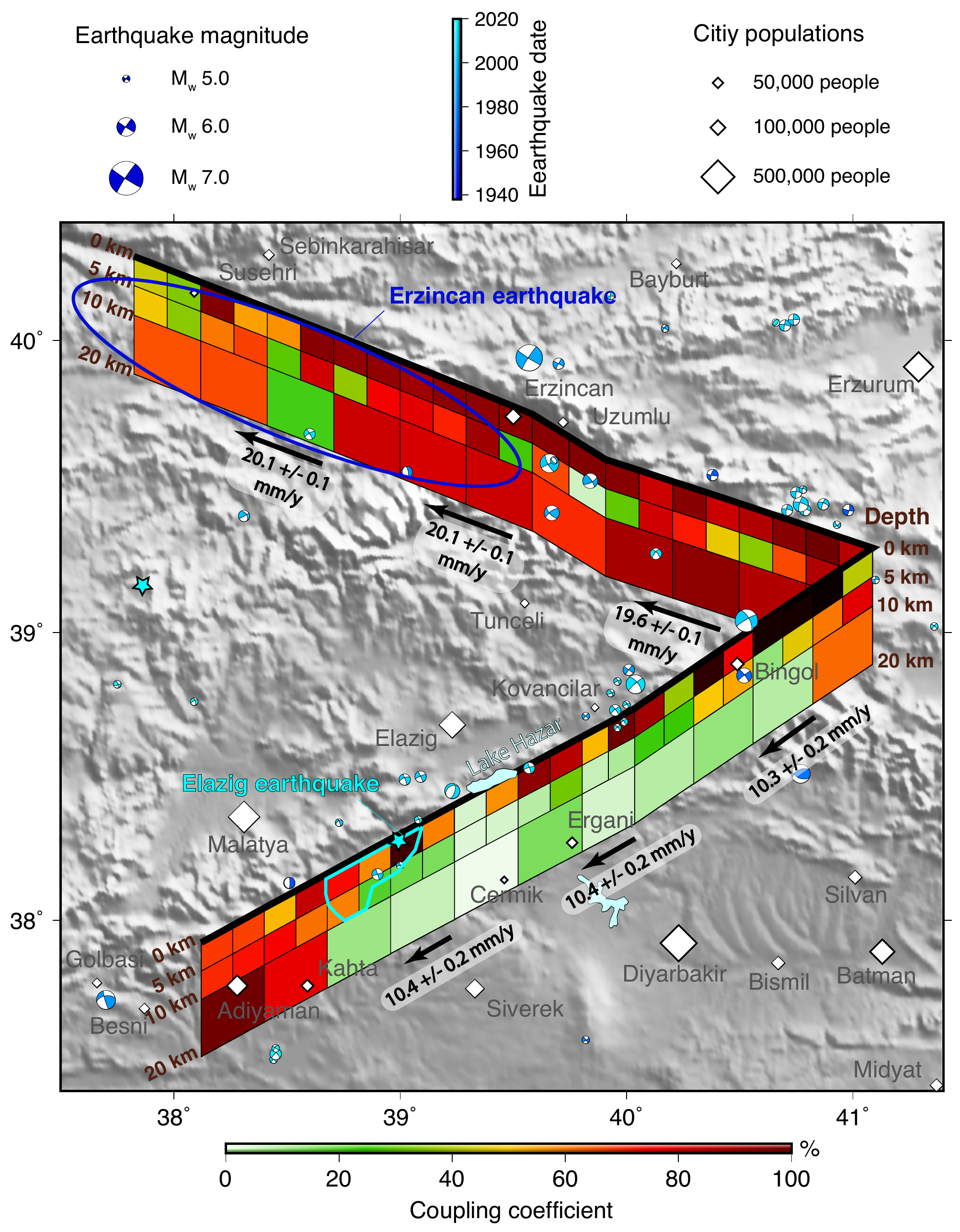}
\vspace{-.75cm}\hspace{-1.5cm}\caption{Interseismic coupling distribution inverted from InSAR and GPS data (mean of posterior pdfs in Figs. S10-S11). Black thick arrows indicate the long-term slip rate at depth derived from the inversion (mean and standard deviation of posterior pdfs in Fig S9). Focal mechanisms show M $>$ 4.8 earthquakes (colors indicate event dates). Contours delineate the approximate rupture extent of the 1939 $\mathrm{M}_S$ 8.0 Erzincan earthquake and of the 2020 $\mathrm{M}_w$ 6.8 Elaz{\i}\u{g} earthquake (USGS finite fault solution). The light blue star indicates the epicenter of the Elaz{\i}\u{g} earthquake.}
\label{coupling}
\end{figure}

For each posterior Euler pole, we calculate the rotation predicted at the center of each patch and project the obtained vector along the fault strike direction to obtain posterior pdfs of the long-term slip rate along the faults (Fig. S9). These pdfs are consistent with steady long term slip rates of $\sim$ 20 mm/y along the NAF and $\sim$ 10 mm/y along the EAF (Figs. \ref{coupling}, S9, Tables S1, S2).  

\begin{figure}
\hspace{-1.5cm}\noindent\includegraphics[width=17cm]{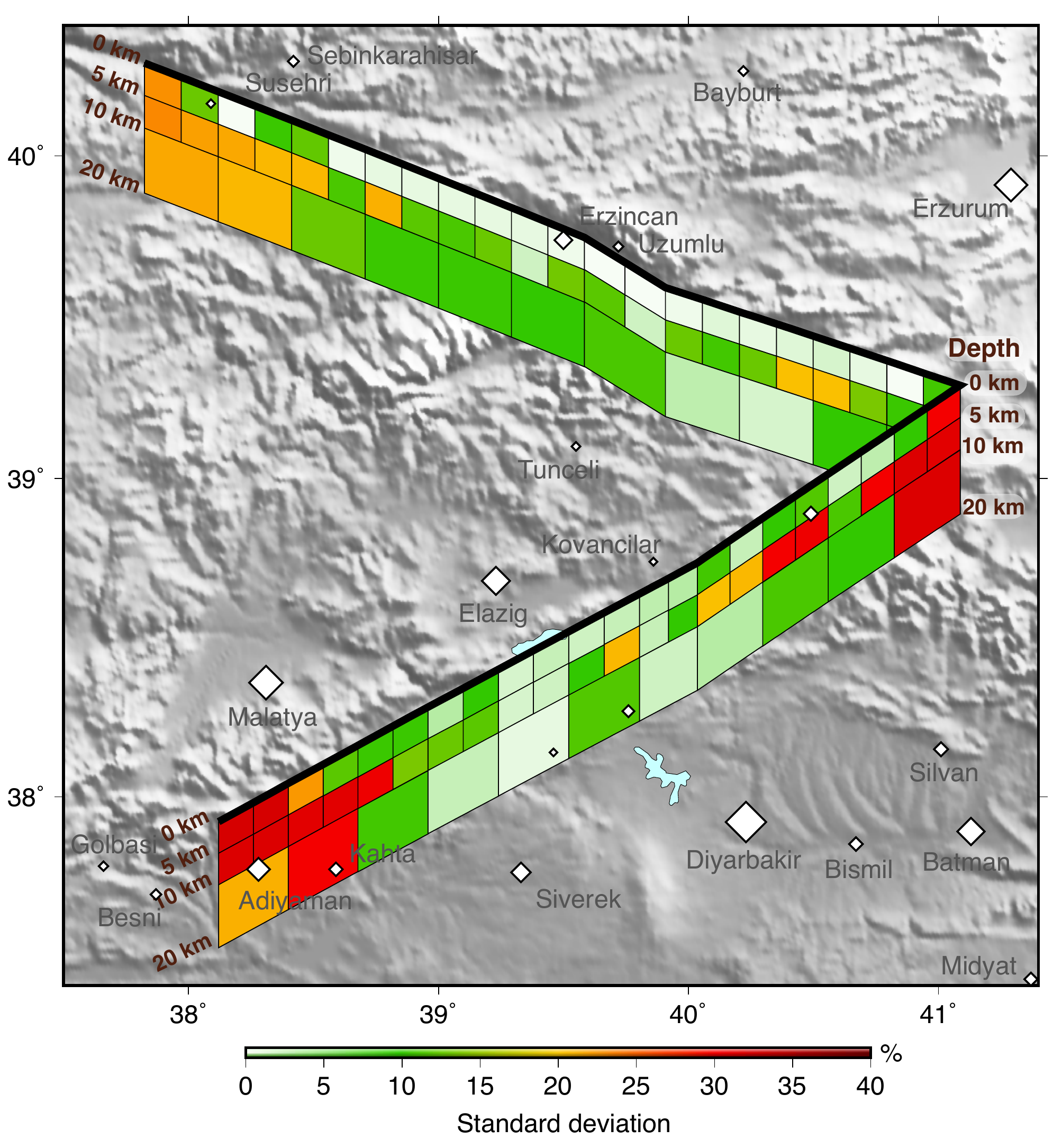}
\caption{Standard deviation of the coupling posterior pdfs. The extreme west and -- to a lesser extent -- the extreme east parts of the NAF-EAF system, presenting high ($>25$ \%) standard deviations, are located outside of the InSAR tracks.}
\label{std}
\end{figure}

For each sampled model $\mathbf{m}_k = (\mathbf{w}^1_k, \mathbf{w}^2_k, \mathbf{S}_k)^T$, we divide the back-slip parameters $\mathbf{S}_k$ by the long-term fault rate calculated at the center of each patch using the corresponding sampled Euler poles $\mathbf{w}^1_k$ and $\mathbf{w}^2_k$ to obtain the posterior marginal pdfs on the coupling coefficients (Figs. S10, S11).
We show these pdfs in the form of their means (Fig. \ref{coupling}) and standard deviations (Fig. \ref{std}). Although restrictive, this representation gives an approximate view of the coupling spatial distribution and its associated uncertainties. Uncertainty is high ($>25$ \%) on the extreme west and -- to a lesser extent -- the extreme east parts of the fault system which are located outside of the InSAR tracks (Fig. \ref{intro}). The standard deviation on most parts of the faults is $<$ 20\%, much lower on many subfaults (Fig. \ref{std}). Note that standard deviation values are likely under-estimated since we did not consider epistemic uncertainties here. The Earth structure is likely not homogeneous and the fault geometry not as simple as we modeled it, generating more uncertainties that we do not account for.

We calculate the GPS and InSAR measurements predicted for every posterior sampled model. We plot the predicted GPS means (red arrows) and 2-$\sigma$ standard deviations (red ellipses) on Fig. S12 and the residuals on Fig. S13. For InSAR, we plot the mean predicted LOS displacements (Figs. S14-S16) and standard deviations (Fig. S17). The range of likely models that we found (Figs. S10-S11) is in very good agreement with both GPS and InSAR data. One way to quantify the relative amplitudes of residuals with respect to the observations is to calculate the ratio $r$ of the mean of the absolute value of the residuals with the mean of the absolute value of the observations,
\begin{linenomath*}
\begin{equation}
r = \frac{<| \mathbf{d} - \mathbf{d_{pred}}  |>}{<| \mathbf{d}  |>}.
\end{equation}
\end{linenomath*}
This ratio is 15.9 \% for T264, 36.1\% for T400, 24.3 \% for T493, 21.6 \% for GPS. We attribute these reasonably small residuals -- which do not exhibit coherent pattern (Figs. S13, S17) -- to non tectonic sources. Furthermore, we find that every posterior sampled model predict very similar GPS and InSAR displacements; red ellipses are hardly visible on Fig. S12 and the standard deviations of the predicted InSAR LOS displacements are very small (Fig. S17). This highlights the limited resolution on the coupling model: if different models predict the same observations, discriminating between them is difficult. 

\section{Discussion}

We show focal mechanisms of $\mathrm{M}>$ 4.8 earthquakes in the studied area from the Global Centroid Moment Tensor (GCMT) catalog \cite{dziewonski1981determination, ekstrom2012global} for events posterior to 1976 and from a compilation of historical earthquakes \cite{tan2008earthquake} for earlier events (1938 -- 1976) (Fig. \ref{coupling}). Focal mechanisms are represented at the location of their surface projections (i.e. at depth $=$ 0). Colors indicate the dates of the events. The largest earthquake in the studied area is the 1939 $\mathrm{M}_S$ 8.0 Erzincan earthquake which initiated near Erzincan and extended over the entire NAF segment west of Erzincan represented in Fig. \ref{coupling} \cite{barka1996slip, stein1997progressive}. We find that almost all of this section is strongly coupled, such as the rest of the studied NAF segment east of Erzincan. This easternmost segment of the NAF presents a moderate seismicity compared to the rest of the NAF. Our interseismic slip distribution suggests that it is as prone to generate large earthquakes as the rest of the NAF and as the Erzincan rupture segment in particular. In the middle of this overall strongly-coupled ($>$ 75\%) fault, we identify a few low-to-moderate coupling (10-50\%) patches at depths between 5 and 10 km (Fig. \ref{coupling}). These patches are associated with standard deviations between 5 and 25 \%, suggesting that these uncoupled patches are robust features. Interestingly, the most uncoupled patch coincides with the main step-over of this section of the NAF. Step-overs are thought to act as geometrical barriers that stop earthquake ruptures \cite <e.g., >[]{wesnousky2006predicting}. Although limited to one example, our results suggest that these geometrical features may also influence -- or be influenced by -- the intereseismic behavior of the faults.

We find that locking on the EAF is much shallower with coupling values $>50$ \% limited to the first 5 km, consistently with previous studies \cite{cavalie2014block}. High coupling found at depth on the westernmost part of the fault is associated with standard deviations $>$ 20 \%, meaning that they are not reliable (Fig. \ref{std}). Furthermore, we find that coupling also varies within the shallowest portion of the fault, alternating strongly coupled segments with weakly-to-moderately (0-60\%) coupled ones (Fig. \ref{coupling}). The most uncoupled shallow fault portion of the central EAF is located near Elaz{\i}\u{g}, and coincides with the pull apart basin of Lake Hazar, as also observed on the Haiyuan fault \cite{jolivet2013spatio}. Different stress orientations around the basin could favor low coupling \cite{bertoluzza1997finite, wang2017strain, van2017initiation}. This large reservoir of water may also provide the shallow part of the fault with fluids \cite<although low resistivity associated to fluids is rather observed below 10 km depth,>{turkouglu2015crustal}, and locally weaken its mechanical friction, favoring asesimic slip. Such a behavior is observed both in laboratory and in situ (at the decametric scale) \cite{cappa2019stabilization}. The mechanism invoked by the authors -- consisting in an increase in nucleation length due to an increase in pore fluid pressure -- may be at play at much larger scale here. On the other hand, the few earthquakes recorded on the EAF coincide with relatively high coupling. Before the recent Elaz{\i}\u{g} earthquake, the two largest events occurred near the localities of Bingol ($\mathrm{M}_w$ 6.3, 2003) and Kovancilar ($\mathrm{M}_w$ 6.1, 2010). The second one was followed by numerous aftershocks with magnitudes up to 5.6. All of these earthquakes occurred on $>$ 65\% coupled fault portions while fault segments with coupling $<$ 50\% do not appear to have hosted M $>$ 4.8 earthquakes. 

According to the USGS finite-fault model \cite{usgs}, the Elaz{\i}\u{g} earthquake initiated between Elaz{\i}\u{g} and Malatya (light blue star in Fig. \ref{coupling}) and propagated unilaterally westward (light blue contour in Fig. \ref{coupling}). The early part coincides with a strongly-locked (coupling coefficient: 100\%) but narrow (13.7 $\times$ 5 km) patch. The rupture seems to have then propagated throughout moderately coupled (coupling coefficient: 50-80\%) fault segments. Although the USGS model is preliminary, its contours correlate fairly well with the coupling distribution, suggesting that the rupture stopped when reaching $<$ 25\% coupled fault portions (Fig. \ref{coupling}).

\begin{figure}
\noindent\includegraphics[width=14cm]{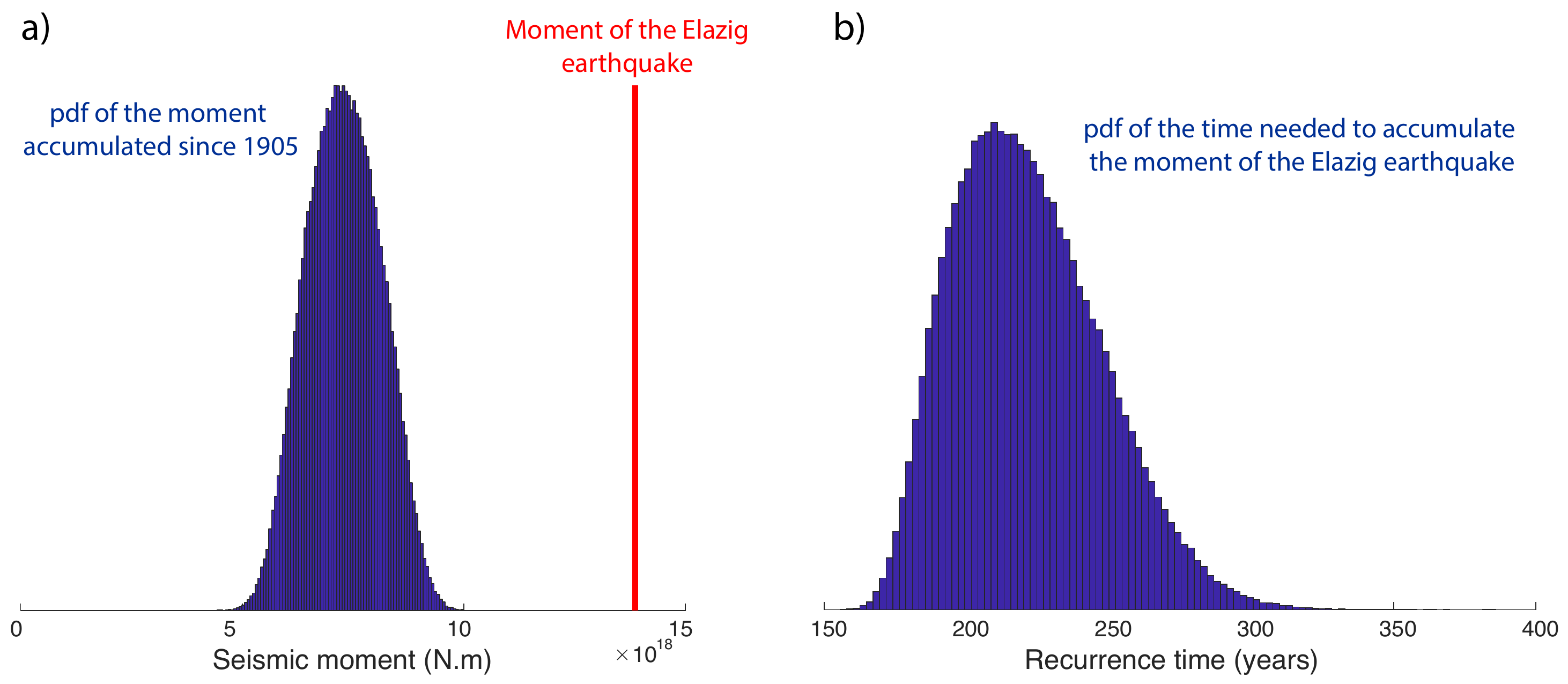}   
\caption{a) Pdf of the accumulated seismic moment on the 4 patches inside the Elaz{\i}\u{g} rupture since 1905. The red vertical line indicates the seismic moment of the Elaz{\i}\u{g} earthquake according to the USGS solution (13.87 $\times 10^{18}$ N.m). b) Pdf of the time necessary to accumulate the seismic moment which was released during the Elaz{\i}\u{g} earthquake.}
\label{Mo}
\end{figure}

The last M $>$ 6.6 earthquake in the approximate region dates back to 1905 ($\mathrm{M}_S =$ 6.7) \cite{nalbant2002stress}. This event was located west of the recent Elaz{\i}\u{g} earthquake (38.6$^{\circ}$ E, 38.1$^{\circ}$ N) \cite{nalbant2002stress} but, given location uncertainties, could have ruptured the same fault portion. We calculate, for each sampled coupling model, the accumulated moment inside the rupture contour of the Elaz{\i}\u{g} earthquake since 1905. To simplify the problem, we assume that the earthquake ruptured the entire surface of the 4 main subfaults inside the rupture contour and not more, i.e. the 3 shallowest subfaults plus the westernmost intermediate-depth one (Fig. 3). We obtain a pdf of the seismic moment accumulated  since 1905 (Fig. \ref{Mo}.a). The pdf mean is 7.3 $\times 10^{18}$ N.m, its standard deviation 0.8 $\times 10^{18}$ N.m. According to the USGS solution, the seismic moment released during the Elaz{\i}\u{g} earthquake is 13.87 $\times 10^{18}$ N.m -- other solutions find even larger seismic moments \cite< e.g., GCMT,>[]{pousse20202020} -- which is much larger than the $7.3 \pm 0.8 \times 10^{18}$ N.m of moment deficit that we estimated since 1905. This seems to indicate that the recent Elaz{\i}\u{g} earthquake did not rupture the same fault portion than the 1905 earthquake. We further calculate the pdf of the time necessary to accumulate the seismic moment which was released during the 2020 Elaz{\i}\u{g} earthquake (Fig. \ref{Mo}.b). The mean and standard deviation of the obtained pdf give a recurrence time for an Elaz{\i}\u{g}-type earthquake of 221.5 $\pm$ 26 years. 

\section{Conclusion}

We inverted InSAR and GPS observations to image interseismic coupling along the North and East Anatolian faults in eastern Turkey. We adopted a Bayesian sampling approach in order to estimate posterior uncertainties on the coupling distributions and on the long term fault rate. We did not impose a pre-calculated plate motion but inverted for the rotation of both the Anatolian and Arabian plates with respect to Eurasia, ensuring that the inferred coupling distribution is not biased in a systematic way by an inaccurate plate motion model. We found that the North Anatolian fault is strongly coupled from 0 to 20 km depth while the East Anatolian fault is weakly coupled for the most part with high ($>50$ \%) coupling values limited to the shallowest part of the fault (0 to 5 km). Furthermore, we find that coupling is heterogeneous within this shallow portion, alternating seemingly creeping sections with strongly locked patches. Comparison between our interseismic coupling distribution and the preliminary finite-fault model of the USGS for the 2020 $\rm{M_w}$ 6.8 Elaz{\i}\u{g} earthquake reveals that this event likely initiated on one of this strongly locked (coupling coefficient: 100\%) fault patch and then propagated into moderately coupled fault segments (coupling coefficient: 50-80\%). Overall, we estimate that the Elaz{\i}\u{g} earthquake released 221.5 ($\pm$ 26) years of accumulated moment, suggesting a recurrence time ranging from 200 to 250 years.

\appendix

\section{Rotation matrix $\mathbf{A}$}

We build the rotation matrix $\mathbf{A}$ so that the motion due to the rotation of both the Anatolian and Arabian plates with respect to Eurasia equals $\mathbf{A} \cdot \mathbf{W}$, where 
\begin{linenomath*}
\begin{equation}
\mathbf{W} =
\begin{pmatrix}
\mathbf{w}^1 \\
\mathbf{w}^2
\end{pmatrix}.
\end{equation}
\end{linenomath*}

Sorting all data points located on the Eurasian plate at the beginning of $\mathbf{d}$, all data points located on the Anatolian plate in the middle and all data points located on the Arabian plate at the end, i.e. writing $\mathbf{d}$ as
\begin{linenomath*}
\begin{equation}
\mathbf{d} = 
\begin{pmatrix}
\mathbf{d}_{0} \\
\mathbf{d}_{1} \\
\mathbf{d}_{2} 
\end{pmatrix},
\end{equation}
\end{linenomath*}
with $\mathbf{d}_{0}$, $\mathbf{d}_{1}$, $\mathbf{d}_{2}$ data points located on the Eurasian, Anatolian and Arabian plates respectively, we can write $\mathbf{A}$ as a block matrix 
\begin{linenomath*}
\begin{equation}
\mathbf{A} = 
\begin{pmatrix}
\mathbf{0}  & \mathbf{0} \\
\mathbf{A'} & \mathbf{0} \\
\mathbf{0}  & \mathbf{A'} 
\end{pmatrix},
\end{equation}
\end{linenomath*}
so that $\mathbf{A} \cdot \mathbf{W}$ equals $\mathbf{0}$ for data points in Eurasia, $\mathbf{A'} \cdot \mathbf{w}^1$ in Anatolia and $\mathbf{A}' \cdot \mathbf{w}^2$ in Arabia. $\mathbf{A'}$ is a transfer matrix relating the rotation vector in Cartesian geocentric coordinates $\mathbf{W}$ to the rotation block motion at each data point. It can be expressed at the location of an InSAR pixel or GPS station of longitude $\lambda$ and latitude $\phi$ as 
\begin{linenomath*}
\begin{equation}
\mathbf{A'}_{\lambda,\phi} =
\begin{pmatrix}
- \sin{\lambda} & \cos{\lambda} & 0 \\
- \sin{\phi} \cos{\lambda} & - \sin{\phi} \sin{\lambda} & \cos{\phi}\\ 
\cos{\phi} \cos{\lambda} & \cos{\phi} \sin{\lambda} & \sin{\phi}
\end{pmatrix}
\cdot
\begin{pmatrix}
0 & z & -y \\
-z & 0 & x \\
y & -x & 0
\end{pmatrix},
\end{equation}
\end{linenomath*}
where
\begin{linenomath*}
\begin{equation}
\begin{pmatrix}
x \\
y \\ 
z
\end{pmatrix}
= R_e (1 - \epsilon \sin^2{\phi})^{-1/2}
\begin{pmatrix}
\cos{\phi} \cos{\lambda} \\
\cos{\phi} \sin{\lambda} \\ 
(1-\epsilon) \sin{\phi}
\end{pmatrix},
\end{equation}
\end{linenomath*}
with $R_e = 6378.137$ km the Earth equatorial radius and $\epsilon = 0.00669438003$ the Earth eccentricity \cite{bowring1985accuracy}.

\acknowledgments
The SAR data were provided by the European Space Agency through category‐1 project 6703, accessible at the following address: https://earth.esa.int/web/guest/data-access/. The inversion was performed using the AlTar software developed by Sarah Minson, Junle Jiang, Hailiang Zhang, Romain Jolivet, Zacharie Duputel, Michael Aivazis, James Beck and Mark Simons at Caltech. We thank the Editor as well as Romain Jolivet and A. Ozgun Konca for thourough reviews.


%
%

\bibliography{anatolia_biblio}

\begin{thebibliography}{}

\bibitem [\protect \citeauthoryear {%
Aktug%
, Dikmen%
, Dogru%
\BCBL {}\ \BBA {} Ozener%
}{%
Aktug%
\ \protect \BOthers {.}}{%
{\protect \APACyear {2013}}%
}]{%
aktug2013seismicity}
\APACinsertmetastar {%
aktug2013seismicity}%
\begin{APACrefauthors}%
Aktug, B.%
, Dikmen, U.%
, Dogru, A.%
\BCBL {}\ \BBA {} Ozener, H.%
\end{APACrefauthors}%
\unskip\
\newblock
\APACrefYearMonthDay{2013}{}{}.
\newblock
{\BBOQ}\APACrefatitle {Seismicity and strain accumulation around Karliova
  triple junction (Turkey)} {Seismicity and strain accumulation around karliova
  triple junction (turkey)}.{\BBCQ}
\newblock
\APACjournalVolNumPages{Journal of Geodynamics}{67}{}{21--29}.
\PrintBackRefs{\CurrentBib}

\bibitem [\protect \citeauthoryear {%
Aktug%
\ \protect \BOthers {.}}{%
Aktug%
\ \protect \BOthers {.}}{%
{\protect \APACyear {2016}}%
}]{%
aktug2016slip}
\APACinsertmetastar {%
aktug2016slip}%
\begin{APACrefauthors}%
Aktug, B.%
, Ozener, H.%
, Dogru, A.%
, Sabuncu, A.%
, Turgut, B.%
, Halicioglu, K.%
\BDBL {}Havazli, E.%
\end{APACrefauthors}%
\unskip\
\newblock
\APACrefYearMonthDay{2016}{}{}.
\newblock
{\BBOQ}\APACrefatitle {Slip rates and seismic potential on the East Anatolian
  Fault System using an improved GPS velocity field} {Slip rates and seismic
  potential on the east anatolian fault system using an improved gps velocity
  field}.{\BBCQ}
\newblock
\APACjournalVolNumPages{Journal of Geodynamics}{94}{}{1--12}.
\PrintBackRefs{\CurrentBib}

\bibitem [\protect \citeauthoryear {%
Ambraseys%
}{%
Ambraseys%
}{%
{\protect \APACyear {1971}}%
}]{%
ambraseys1971value}
\APACinsertmetastar {%
ambraseys1971value}%
\begin{APACrefauthors}%
Ambraseys, N.%
\end{APACrefauthors}%
\unskip\
\newblock
\APACrefYearMonthDay{1971}{}{}.
\newblock
{\BBOQ}\APACrefatitle {Value of historical records of earthquakes} {Value of
  historical records of earthquakes}.{\BBCQ}
\newblock
\APACjournalVolNumPages{Nature}{232}{5310}{375--379}.
\PrintBackRefs{\CurrentBib}

\bibitem [\protect \citeauthoryear {%
Ambraseys%
}{%
Ambraseys%
}{%
{\protect \APACyear {1989}}%
}]{%
ambraseys1989temporary}
\APACinsertmetastar {%
ambraseys1989temporary}%
\begin{APACrefauthors}%
Ambraseys, N.%
\end{APACrefauthors}%
\unskip\
\newblock
\APACrefYearMonthDay{1989}{}{}.
\newblock
{\BBOQ}\APACrefatitle {Temporary seismic quiescence: SE Turkey} {Temporary
  seismic quiescence: Se turkey}.{\BBCQ}
\newblock
\APACjournalVolNumPages{Geophysical Journal International}{96}{2}{311--331}.
\PrintBackRefs{\CurrentBib}

\bibitem [\protect \citeauthoryear {%
Avouac%
}{%
Avouac%
}{%
{\protect \APACyear {2015}}%
}]{%
avouac2015geodetic}
\APACinsertmetastar {%
avouac2015geodetic}%
\begin{APACrefauthors}%
Avouac, J\BHBI P.%
\end{APACrefauthors}%
\unskip\
\newblock
\APACrefYearMonthDay{2015}{}{}.
\newblock
{\BBOQ}\APACrefatitle {From geodetic imaging of seismic and aseismic fault slip
  to dynamic modeling of the seismic cycle} {From geodetic imaging of seismic
  and aseismic fault slip to dynamic modeling of the seismic cycle}.{\BBCQ}
\newblock
\APACjournalVolNumPages{Annual Review of Earth and Planetary
  Sciences}{43}{}{233--271}.
\PrintBackRefs{\CurrentBib}

\bibitem [\protect \citeauthoryear {%
Barka%
}{%
Barka%
}{%
{\protect \APACyear {1996}}%
}]{%
barka1996slip}
\APACinsertmetastar {%
barka1996slip}%
\begin{APACrefauthors}%
Barka, A.%
\end{APACrefauthors}%
\unskip\
\newblock
\APACrefYearMonthDay{1996}{}{}.
\newblock
{\BBOQ}\APACrefatitle {Slip distribution along the North Anatolian fault
  associated with the large earthquakes of the period 1939 to 1967} {Slip
  distribution along the north anatolian fault associated with the large
  earthquakes of the period 1939 to 1967}.{\BBCQ}
\newblock
\APACjournalVolNumPages{Bulletin of the Seismological Society of
  America}{86}{5}{1238--1254}.
\PrintBackRefs{\CurrentBib}

\bibitem [\protect \citeauthoryear {%
Bertoluzza%
\ \BBA {} Perotti%
}{%
Bertoluzza%
\ \BBA {} Perotti%
}{%
{\protect \APACyear {1997}}%
}]{%
bertoluzza1997finite}
\APACinsertmetastar {%
bertoluzza1997finite}%
\begin{APACrefauthors}%
Bertoluzza, L.%
\BCBT {}\ \BBA {} Perotti, C\BPBI R.%
\end{APACrefauthors}%
\unskip\
\newblock
\APACrefYearMonthDay{1997}{}{}.
\newblock
{\BBOQ}\APACrefatitle {A finite-element model of the stress field in
  strike-slip basins: implications for the Permian tectonics of the Southern
  Alps (Italy)} {A finite-element model of the stress field in strike-slip
  basins: implications for the permian tectonics of the southern alps
  (italy)}.{\BBCQ}
\newblock
\APACjournalVolNumPages{Tectonophysics}{280}{1-2}{185--197}.
\PrintBackRefs{\CurrentBib}

\bibitem [\protect \citeauthoryear {%
Bletery%
, Sladen%
, Jiang%
\BCBL {}\ \BBA {} Simons%
}{%
Bletery%
\ \protect \BOthers {.}}{%
{\protect \APACyear {2016}}%
}]{%
bletery2016bayesian}
\APACinsertmetastar {%
bletery2016bayesian}%
\begin{APACrefauthors}%
Bletery, Q.%
, Sladen, A.%
, Jiang, J.%
\BCBL {}\ \BBA {} Simons, M.%
\end{APACrefauthors}%
\unskip\
\newblock
\APACrefYearMonthDay{2016}{}{}.
\newblock
{\BBOQ}\APACrefatitle {A Bayesian source model for the 2004 great
  Sumatra-Andaman earthquake} {A bayesian source model for the 2004 great
  sumatra-andaman earthquake}.{\BBCQ}
\newblock
\APACjournalVolNumPages{Journal of Geophysical Research: Solid
  Earth}{121}{7}{5116--5135}.
\PrintBackRefs{\CurrentBib}

\bibitem [\protect \citeauthoryear {%
Bowring%
}{%
Bowring%
}{%
{\protect \APACyear {1985}}%
}]{%
bowring1985accuracy}
\APACinsertmetastar {%
bowring1985accuracy}%
\begin{APACrefauthors}%
Bowring, B.%
\end{APACrefauthors}%
\unskip\
\newblock
\APACrefYearMonthDay{1985}{}{}.
\newblock
{\BBOQ}\APACrefatitle {The accuracy of geodetic latitude and height equations}
  {The accuracy of geodetic latitude and height equations}.{\BBCQ}
\newblock
\APACjournalVolNumPages{Survey Review}{28}{218}{202--206}.
\PrintBackRefs{\CurrentBib}

\bibitem [\protect \citeauthoryear {%
B{\"u}rgmann%
\ \protect \BOthers {.}}{%
B{\"u}rgmann%
\ \protect \BOthers {.}}{%
{\protect \APACyear {2005}}%
}]{%
burgmann2005interseismic}
\APACinsertmetastar {%
burgmann2005interseismic}%
\begin{APACrefauthors}%
B{\"u}rgmann, R.%
, Kogan, M\BPBI G.%
, Steblov, G\BPBI M.%
, Hilley, G.%
, Levin, V\BPBI E.%
\BCBL {}\ \BBA {} Apel, E.%
\end{APACrefauthors}%
\unskip\
\newblock
\APACrefYearMonthDay{2005}{}{}.
\newblock
{\BBOQ}\APACrefatitle {Interseismic coupling and asperity distribution along
  the Kamchatka subduction zone} {Interseismic coupling and asperity
  distribution along the kamchatka subduction zone}.{\BBCQ}
\newblock
\APACjournalVolNumPages{Journal of Geophysical Research: Solid
  Earth}{110}{B7}{}.
\PrintBackRefs{\CurrentBib}

\bibitem [\protect \citeauthoryear {%
Burton%
, McGonigle%
, Makropoulos%
\BCBL {}\ \BBA {} {\"U}{\c{c}}er%
}{%
Burton%
\ \protect \BOthers {.}}{%
{\protect \APACyear {1984}}%
}]{%
burton1984seismic}
\APACinsertmetastar {%
burton1984seismic}%
\begin{APACrefauthors}%
Burton, P\BPBI W.%
, McGonigle, R.%
, Makropoulos, K\BPBI C.%
\BCBL {}\ \BBA {} {\"U}{\c{c}}er, S\BPBI B.%
\end{APACrefauthors}%
\unskip\
\newblock
\APACrefYearMonthDay{1984}{}{}.
\newblock
{\BBOQ}\APACrefatitle {Seismic risk in Turkey, the Aegean and the eastern
  Mediterranean: the occurrence of large magnitude earthquakes} {Seismic risk
  in turkey, the aegean and the eastern mediterranean: the occurrence of large
  magnitude earthquakes}.{\BBCQ}
\newblock
\APACjournalVolNumPages{Geophysical Journal International}{78}{2}{475--506}.
\PrintBackRefs{\CurrentBib}

\bibitem [\protect \citeauthoryear {%
Cappa%
, Scuderi%
, Collettini%
, Guglielmi%
\BCBL {}\ \BBA {} Avouac%
}{%
Cappa%
\ \protect \BOthers {.}}{%
{\protect \APACyear {2019}}%
}]{%
cappa2019stabilization}
\APACinsertmetastar {%
cappa2019stabilization}%
\begin{APACrefauthors}%
Cappa, F.%
, Scuderi, M\BPBI M.%
, Collettini, C.%
, Guglielmi, Y.%
\BCBL {}\ \BBA {} Avouac, J\BHBI P.%
\end{APACrefauthors}%
\unskip\
\newblock
\APACrefYearMonthDay{2019}{}{}.
\newblock
{\BBOQ}\APACrefatitle {Stabilization of fault slip by fluid injection in the
  laboratory and in situ} {Stabilization of fault slip by fluid injection in
  the laboratory and in situ}.{\BBCQ}
\newblock
\APACjournalVolNumPages{Science advances}{5}{3}{eaau4065}.
\PrintBackRefs{\CurrentBib}

\bibitem [\protect \citeauthoryear {%
Cavali{\'e}%
\ \BBA {} J{\'o}nsson%
}{%
Cavali{\'e}%
\ \BBA {} J{\'o}nsson%
}{%
{\protect \APACyear {2014}}%
}]{%
cavalie2014block}
\APACinsertmetastar {%
cavalie2014block}%
\begin{APACrefauthors}%
Cavali{\'e}, O.%
\BCBT {}\ \BBA {} J{\'o}nsson, S.%
\end{APACrefauthors}%
\unskip\
\newblock
\APACrefYearMonthDay{2014}{}{}.
\newblock
{\BBOQ}\APACrefatitle {Block-like plate movements in eastern Anatolia observed
  by InSAR} {Block-like plate movements in eastern anatolia observed by
  insar}.{\BBCQ}
\newblock
\APACjournalVolNumPages{Geophysical Research Letters}{41}{1}{26--31}.
\PrintBackRefs{\CurrentBib}

\bibitem [\protect \citeauthoryear {%
Doin%
\ \protect \BOthers {.}}{%
Doin%
\ \protect \BOthers {.}}{%
{\protect \APACyear {2011}}%
}]{%
doin2011presentation}
\APACinsertmetastar {%
doin2011presentation}%
\begin{APACrefauthors}%
Doin, M\BHBI P.%
, Guillaso, S.%
, Jolivet, R.%
, Lasserre, C.%
, Lodge, F.%
, Ducret, G.%
\BCBL {}\ \BBA {} Grandin, R.%
\end{APACrefauthors}%
\unskip\
\newblock
\APACrefYearMonthDay{2011}{}{}.
\newblock
{\BBOQ}\APACrefatitle {Presentation of the small baseline NSBAS processing
  chain on a case example: the Etna deformation monitoring from 2003 to 2010
  using Envisat data} {Presentation of the small baseline nsbas processing
  chain on a case example: the etna deformation monitoring from 2003 to 2010
  using envisat data}.{\BBCQ}
\newblock
\BIn{} \APACrefbtitle {Proceedings of the Fringe symposium} {Proceedings of the
  fringe symposium}\ (\BPGS\ 3434--3437).
\PrintBackRefs{\CurrentBib}

\bibitem [\protect \citeauthoryear {%
Duman%
\ \BBA {} Emre%
}{%
Duman%
\ \BBA {} Emre%
}{%
{\protect \APACyear {2013}}%
}]{%
duman2013east}
\APACinsertmetastar {%
duman2013east}%
\begin{APACrefauthors}%
Duman, T\BPBI Y.%
\BCBT {}\ \BBA {} Emre, {\"O}.%
\end{APACrefauthors}%
\unskip\
\newblock
\APACrefYearMonthDay{2013}{}{}.
\newblock
{\BBOQ}\APACrefatitle {The East Anatolian Fault: geometry, segmentation and jog
  characteristics} {The east anatolian fault: geometry, segmentation and jog
  characteristics}.{\BBCQ}
\newblock
\APACjournalVolNumPages{Geological Society, London, Special
  Publications}{372}{1}{495--529}.
\PrintBackRefs{\CurrentBib}

\bibitem [\protect \citeauthoryear {%
Duputel%
, Agram%
, Simons%
, Minson%
\BCBL {}\ \BBA {} Beck%
}{%
Duputel%
\ \protect \BOthers {.}}{%
{\protect \APACyear {2014}}%
}]{%
duputel2014accounting}
\APACinsertmetastar {%
duputel2014accounting}%
\begin{APACrefauthors}%
Duputel, Z.%
, Agram, P\BPBI S.%
, Simons, M.%
, Minson, S\BPBI E.%
\BCBL {}\ \BBA {} Beck, J\BPBI L.%
\end{APACrefauthors}%
\unskip\
\newblock
\APACrefYearMonthDay{2014}{}{}.
\newblock
{\BBOQ}\APACrefatitle {Accounting for prediction uncertainty when inferring
  subsurface fault slip} {Accounting for prediction uncertainty when inferring
  subsurface fault slip}.{\BBCQ}
\newblock
\APACjournalVolNumPages{Geophysical Journal International}{197}{1}{464--482}.
\PrintBackRefs{\CurrentBib}

\bibitem [\protect \citeauthoryear {%
Dziewonski%
, Chou%
\BCBL {}\ \BBA {} Woodhouse%
}{%
Dziewonski%
\ \protect \BOthers {.}}{%
{\protect \APACyear {1981}}%
}]{%
dziewonski1981determination}
\APACinsertmetastar {%
dziewonski1981determination}%
\begin{APACrefauthors}%
Dziewonski, A.%
, Chou, T\BHBI A.%
\BCBL {}\ \BBA {} Woodhouse, J.%
\end{APACrefauthors}%
\unskip\
\newblock
\APACrefYearMonthDay{1981}{}{}.
\newblock
{\BBOQ}\APACrefatitle {Determination of earthquake source parameters from
  waveform data for studies of global and regional seismicity} {Determination
  of earthquake source parameters from waveform data for studies of global and
  regional seismicity}.{\BBCQ}
\newblock
\APACjournalVolNumPages{Journal of Geophysical Research: Solid
  Earth}{86}{B4}{2825--2852}.
\PrintBackRefs{\CurrentBib}

\bibitem [\protect \citeauthoryear {%
Ekstr{\"o}m%
, Nettles%
\BCBL {}\ \BBA {} Dziewo{\'n}ski%
}{%
Ekstr{\"o}m%
\ \protect \BOthers {.}}{%
{\protect \APACyear {2012}}%
}]{%
ekstrom2012global}
\APACinsertmetastar {%
ekstrom2012global}%
\begin{APACrefauthors}%
Ekstr{\"o}m, G.%
, Nettles, M.%
\BCBL {}\ \BBA {} Dziewo{\'n}ski, A.%
\end{APACrefauthors}%
\unskip\
\newblock
\APACrefYearMonthDay{2012}{}{}.
\newblock
{\BBOQ}\APACrefatitle {The global CMT project 2004--2010: Centroid-moment
  tensors for 13,017 earthquakes} {The global cmt project 2004--2010:
  Centroid-moment tensors for 13,017 earthquakes}.{\BBCQ}
\newblock
\APACjournalVolNumPages{Physics of the Earth and Planetary
  Interiors}{200}{}{1--9}.
\PrintBackRefs{\CurrentBib}

\bibitem [\protect \citeauthoryear {%
England%
, Houseman%
\BCBL {}\ \BBA {} Nocquet%
}{%
England%
\ \protect \BOthers {.}}{%
{\protect \APACyear {2016}}%
}]{%
england2016constraints}
\APACinsertmetastar {%
england2016constraints}%
\begin{APACrefauthors}%
England, P.%
, Houseman, G.%
\BCBL {}\ \BBA {} Nocquet, J\BHBI M.%
\end{APACrefauthors}%
\unskip\
\newblock
\APACrefYearMonthDay{2016}{}{}.
\newblock
{\BBOQ}\APACrefatitle {Constraints from GPS measurements on the dynamics of
  deformation in Anatolia and the Aegean} {Constraints from gps measurements on
  the dynamics of deformation in anatolia and the aegean}.{\BBCQ}
\newblock
\APACjournalVolNumPages{Journal of Geophysical Research: Solid
  Earth}{121}{12}{8888--8916}.
\PrintBackRefs{\CurrentBib}

\bibitem [\protect \citeauthoryear {%
Jackson%
\ \BBA {} McKenzie%
}{%
Jackson%
\ \BBA {} McKenzie%
}{%
{\protect \APACyear {1988}}%
}]{%
jackson1988relationship}
\APACinsertmetastar {%
jackson1988relationship}%
\begin{APACrefauthors}%
Jackson, J.%
\BCBT {}\ \BBA {} McKenzie, D.%
\end{APACrefauthors}%
\unskip\
\newblock
\APACrefYearMonthDay{1988}{}{}.
\newblock
{\BBOQ}\APACrefatitle {The relationship between plate motions and seismic
  moment tensors, and the rates of active deformation in the Mediterranean and
  Middle East} {The relationship between plate motions and seismic moment
  tensors, and the rates of active deformation in the mediterranean and middle
  east}.{\BBCQ}
\newblock
\APACjournalVolNumPages{Geophysical Journal International}{93}{1}{45--73}.
\PrintBackRefs{\CurrentBib}

\bibitem [\protect \citeauthoryear {%
Jolivet%
\ \protect \BOthers {.}}{%
Jolivet%
\ \protect \BOthers {.}}{%
{\protect \APACyear {2012}}%
}]{%
jolivet2012shallow}
\APACinsertmetastar {%
jolivet2012shallow}%
\begin{APACrefauthors}%
Jolivet, R.%
, Lasserre, C.%
, Doin, M\BHBI P.%
, Guillaso, S.%
, Peltzer, G.%
, Dailu, R.%
\BDBL {}Xu, X.%
\end{APACrefauthors}%
\unskip\
\newblock
\APACrefYearMonthDay{2012}{}{}.
\newblock
{\BBOQ}\APACrefatitle {Shallow creep on the Haiyuan fault (Gansu, China)
  revealed by SAR interferometry} {Shallow creep on the haiyuan fault (gansu,
  china) revealed by sar interferometry}.{\BBCQ}
\newblock
\APACjournalVolNumPages{Journal of Geophysical Research: Solid
  Earth}{117}{B6}{}.
\PrintBackRefs{\CurrentBib}

\bibitem [\protect \citeauthoryear {%
Jolivet%
\ \protect \BOthers {.}}{%
Jolivet%
\ \protect \BOthers {.}}{%
{\protect \APACyear {2013}}%
}]{%
jolivet2013spatio}
\APACinsertmetastar {%
jolivet2013spatio}%
\begin{APACrefauthors}%
Jolivet, R.%
, Lasserre, C.%
, Doin, M\BHBI P.%
, Peltzer, G.%
, Avouac, J\BHBI P.%
, Sun, J.%
\BCBL {}\ \BBA {} Dailu, R.%
\end{APACrefauthors}%
\unskip\
\newblock
\APACrefYearMonthDay{2013}{}{}.
\newblock
{\BBOQ}\APACrefatitle {Spatio-temporal evolution of aseismic slip along the
  Haiyuan fault, China: Implications for fault frictional properties}
  {Spatio-temporal evolution of aseismic slip along the haiyuan fault, china:
  Implications for fault frictional properties}.{\BBCQ}
\newblock
\APACjournalVolNumPages{Earth and Planetary Science Letters}{377}{}{23--33}.
\PrintBackRefs{\CurrentBib}

\bibitem [\protect \citeauthoryear {%
Jolivet%
, Simons%
, Agram%
, Duputel%
\BCBL {}\ \BBA {} Shen%
}{%
Jolivet%
\ \protect \BOthers {.}}{%
{\protect \APACyear {2015}}%
}]{%
jolivet2015aseismic}
\APACinsertmetastar {%
jolivet2015aseismic}%
\begin{APACrefauthors}%
Jolivet, R.%
, Simons, M.%
, Agram, P.%
, Duputel, Z.%
\BCBL {}\ \BBA {} Shen, Z\BHBI K.%
\end{APACrefauthors}%
\unskip\
\newblock
\APACrefYearMonthDay{2015}{}{}.
\newblock
{\BBOQ}\APACrefatitle {Aseismic slip and seismogenic coupling along the central
  San Andreas Fault} {Aseismic slip and seismogenic coupling along the central
  san andreas fault}.{\BBCQ}
\newblock
\APACjournalVolNumPages{Geophysical Research Letters}{42}{2}{297--306}.
\PrintBackRefs{\CurrentBib}

\bibitem [\protect \citeauthoryear {%
Jolivet%
\ \protect \BOthers {.}}{%
Jolivet%
\ \protect \BOthers {.}}{%
{\protect \APACyear {2020}}%
}]{%
jolivet2020interseismic}
\APACinsertmetastar {%
jolivet2020interseismic}%
\begin{APACrefauthors}%
Jolivet, R.%
, Simons, M.%
, Duputel, Z.%
, Olive, J\BHBI A.%
, Bhat, H.%
\BCBL {}\ \BBA {} Bletery, Q.%
\end{APACrefauthors}%
\unskip\
\newblock
\APACrefYearMonthDay{2020}{}{}.
\newblock
{\BBOQ}\APACrefatitle {Interseismic loading of subduction megathrust drives
  long term uplift in northern Chile} {Interseismic loading of subduction
  megathrust drives long term uplift in northern chile}.{\BBCQ}
\newblock
\APACjournalVolNumPages{Geophysical Research Letters}{}{}{}.
\PrintBackRefs{\CurrentBib}

\bibitem [\protect \citeauthoryear {%
J{\'o}nsson%
, Zebker%
, Segall%
\BCBL {}\ \BBA {} Amelung%
}{%
J{\'o}nsson%
\ \protect \BOthers {.}}{%
{\protect \APACyear {2002}}%
}]{%
jonsson2002fault}
\APACinsertmetastar {%
jonsson2002fault}%
\begin{APACrefauthors}%
J{\'o}nsson, S.%
, Zebker, H.%
, Segall, P.%
\BCBL {}\ \BBA {} Amelung, F.%
\end{APACrefauthors}%
\unskip\
\newblock
\APACrefYearMonthDay{2002}{}{}.
\newblock
{\BBOQ}\APACrefatitle {Fault slip distribution of the 1999 M w 7.1 Hector Mine,
  California, earthquake, estimated from satellite radar and GPS measurements}
  {Fault slip distribution of the 1999 m w 7.1 hector mine, california,
  earthquake, estimated from satellite radar and gps measurements}.{\BBCQ}
\newblock
\APACjournalVolNumPages{Bulletin of the Seismological Society of
  America}{92}{4}{1377--1389}.
\PrintBackRefs{\CurrentBib}

\bibitem [\protect \citeauthoryear {%
Kaneko%
, Avouac%
\BCBL {}\ \BBA {} Lapusta%
}{%
Kaneko%
\ \protect \BOthers {.}}{%
{\protect \APACyear {2010}}%
}]{%
kaneko2010towards}
\APACinsertmetastar {%
kaneko2010towards}%
\begin{APACrefauthors}%
Kaneko, Y.%
, Avouac, J\BHBI P.%
\BCBL {}\ \BBA {} Lapusta, N.%
\end{APACrefauthors}%
\unskip\
\newblock
\APACrefYearMonthDay{2010}{}{}.
\newblock
{\BBOQ}\APACrefatitle {Towards inferring earthquake patterns from geodetic
  observations of interseismic coupling} {Towards inferring earthquake patterns
  from geodetic observations of interseismic coupling}.{\BBCQ}
\newblock
\APACjournalVolNumPages{Nature Geoscience}{3}{5}{363--369}.
\PrintBackRefs{\CurrentBib}

\bibitem [\protect \citeauthoryear {%
Le~Pichon%
\ \BBA {} Kreemer%
}{%
Le~Pichon%
\ \BBA {} Kreemer%
}{%
{\protect \APACyear {2010}}%
}]{%
le2010miocene}
\APACinsertmetastar {%
le2010miocene}%
\begin{APACrefauthors}%
Le~Pichon, X.%
\BCBT {}\ \BBA {} Kreemer, C.%
\end{APACrefauthors}%
\unskip\
\newblock
\APACrefYearMonthDay{2010}{}{}.
\newblock
{\BBOQ}\APACrefatitle {The Miocene-to-present kinematic evolution of the
  Eastern Mediterranean and Middle East and its implications for dynamics} {The
  miocene-to-present kinematic evolution of the eastern mediterranean and
  middle east and its implications for dynamics}.{\BBCQ}
\newblock
\APACjournalVolNumPages{Annual Review of Earth and Planetary
  Sciences}{38}{}{323--351}.
\PrintBackRefs{\CurrentBib}

\bibitem [\protect \citeauthoryear {%
Loveless%
\ \BBA {} Meade%
}{%
Loveless%
\ \BBA {} Meade%
}{%
{\protect \APACyear {2011}}%
}]{%
loveless2011spatial}
\APACinsertmetastar {%
loveless2011spatial}%
\begin{APACrefauthors}%
Loveless, J\BPBI P.%
\BCBT {}\ \BBA {} Meade, B\BPBI J.%
\end{APACrefauthors}%
\unskip\
\newblock
\APACrefYearMonthDay{2011}{}{}.
\newblock
{\BBOQ}\APACrefatitle {Spatial correlation of interseismic coupling and
  coseismic rupture extent of the 2011 Mw= 9.0 Tohoku-oki earthquake} {Spatial
  correlation of interseismic coupling and coseismic rupture extent of the 2011
  mw= 9.0 tohoku-oki earthquake}.{\BBCQ}
\newblock
\APACjournalVolNumPages{Geophysical Research Letters}{38}{17}{}.
\PrintBackRefs{\CurrentBib}

\bibitem [\protect \citeauthoryear {%
Mahmoud%
\ \protect \BOthers {.}}{%
Mahmoud%
\ \protect \BOthers {.}}{%
{\protect \APACyear {2013}}%
}]{%
mahmoud2013kinematic}
\APACinsertmetastar {%
mahmoud2013kinematic}%
\begin{APACrefauthors}%
Mahmoud, Y.%
, Masson, F.%
, Meghraoui, M.%
, Cakir, Z.%
, Alchalbi, A.%
, Yavasoglu, H.%
\BDBL {}Inan, S.%
\end{APACrefauthors}%
\unskip\
\newblock
\APACrefYearMonthDay{2013}{}{}.
\newblock
{\BBOQ}\APACrefatitle {Kinematic study at the junction of the East Anatolian
  fault and the Dead Sea fault from GPS measurements} {Kinematic study at the
  junction of the east anatolian fault and the dead sea fault from gps
  measurements}.{\BBCQ}
\newblock
\APACjournalVolNumPages{Journal of Geodynamics}{67}{}{30--39}.
\PrintBackRefs{\CurrentBib}

\bibitem [\protect \citeauthoryear {%
Mansinha%
\ \BBA {} Smylie%
}{%
Mansinha%
\ \BBA {} Smylie%
}{%
{\protect \APACyear {1971}}%
}]{%
mansinha1971displacement}
\APACinsertmetastar {%
mansinha1971displacement}%
\begin{APACrefauthors}%
Mansinha, L\BPBI a.%
\BCBT {}\ \BBA {} Smylie, D.%
\end{APACrefauthors}%
\unskip\
\newblock
\APACrefYearMonthDay{1971}{}{}.
\newblock
{\BBOQ}\APACrefatitle {The displacement fields of inclined faults} {The
  displacement fields of inclined faults}.{\BBCQ}
\newblock
\APACjournalVolNumPages{Bulletin of the Seismological Society of
  America}{61}{5}{1433--1440}.
\PrintBackRefs{\CurrentBib}

\bibitem [\protect \citeauthoryear {%
Marinkovic%
\ \BBA {} Larsen%
}{%
Marinkovic%
\ \BBA {} Larsen%
}{%
{\protect \APACyear {2013}}%
}]{%
marinkovic2013consequences}
\APACinsertmetastar {%
marinkovic2013consequences}%
\begin{APACrefauthors}%
Marinkovic, P.%
\BCBT {}\ \BBA {} Larsen, Y.%
\end{APACrefauthors}%
\unskip\
\newblock
\APACrefYearMonthDay{2013}{}{}.
\newblock
{\BBOQ}\APACrefatitle {Consequences of long-term ASAR local oscillator
  frequency decay—An empirical study of 10 years of data} {Consequences of
  long-term asar local oscillator frequency decay—an empirical study of 10
  years of data}.{\BBCQ}
\newblock
\BIn{} \APACrefbtitle {Living Planet Symposium.} {Living planet symposium.}
\PrintBackRefs{\CurrentBib}

\bibitem [\protect \citeauthoryear {%
Maurer%
\ \BBA {} Johnson%
}{%
Maurer%
\ \BBA {} Johnson%
}{%
{\protect \APACyear {2014}}%
}]{%
maurer2014fault}
\APACinsertmetastar {%
maurer2014fault}%
\begin{APACrefauthors}%
Maurer, J.%
\BCBT {}\ \BBA {} Johnson, K.%
\end{APACrefauthors}%
\unskip\
\newblock
\APACrefYearMonthDay{2014}{}{}.
\newblock
{\BBOQ}\APACrefatitle {Fault coupling and potential for earthquakes on the
  creeping section of the central San Andreas Fault} {Fault coupling and
  potential for earthquakes on the creeping section of the central san andreas
  fault}.{\BBCQ}
\newblock
\APACjournalVolNumPages{Journal of Geophysical Research: Solid
  Earth}{119}{5}{4414--4428}.
\PrintBackRefs{\CurrentBib}

\bibitem [\protect \citeauthoryear {%
Meade%
\ \BBA {} Loveless%
}{%
Meade%
\ \BBA {} Loveless%
}{%
{\protect \APACyear {2009}}%
}]{%
meade2009block}
\APACinsertmetastar {%
meade2009block}%
\begin{APACrefauthors}%
Meade, B\BPBI J.%
\BCBT {}\ \BBA {} Loveless, J\BPBI P.%
\end{APACrefauthors}%
\unskip\
\newblock
\APACrefYearMonthDay{2009}{}{}.
\newblock
{\BBOQ}\APACrefatitle {Block modeling with connected fault-network geometries
  and a linear elastic coupling estimator in spherical coordinates} {Block
  modeling with connected fault-network geometries and a linear elastic
  coupling estimator in spherical coordinates}.{\BBCQ}
\newblock
\APACjournalVolNumPages{Bulletin of the Seismological Society of
  America}{99}{6}{3124--3139}.
\PrintBackRefs{\CurrentBib}

\bibitem [\protect \citeauthoryear {%
Metois%
, Vigny%
\BCBL {}\ \BBA {} Socquet%
}{%
Metois%
\ \protect \BOthers {.}}{%
{\protect \APACyear {2016}}%
}]{%
metois2016interseismic}
\APACinsertmetastar {%
metois2016interseismic}%
\begin{APACrefauthors}%
Metois, M.%
, Vigny, C.%
\BCBL {}\ \BBA {} Socquet, A.%
\end{APACrefauthors}%
\unskip\
\newblock
\APACrefYearMonthDay{2016}{}{}.
\newblock
{\BBOQ}\APACrefatitle {Interseismic coupling, megathrust earthquakes and
  seismic swarms along the Chilean subduction zone (38--18 S)} {Interseismic
  coupling, megathrust earthquakes and seismic swarms along the chilean
  subduction zone (38--18 s)}.{\BBCQ}
\newblock
\APACjournalVolNumPages{Pure and Applied Geophysics}{173}{5}{1431--1449}.
\PrintBackRefs{\CurrentBib}

\bibitem [\protect \citeauthoryear {%
Minson%
, Simons%
\BCBL {}\ \BBA {} Beck%
}{%
Minson%
\ \protect \BOthers {.}}{%
{\protect \APACyear {2013}}%
}]{%
minson2013bayesian}
\APACinsertmetastar {%
minson2013bayesian}%
\begin{APACrefauthors}%
Minson, S.%
, Simons, M.%
\BCBL {}\ \BBA {} Beck, J.%
\end{APACrefauthors}%
\unskip\
\newblock
\APACrefYearMonthDay{2013}{}{}.
\newblock
{\BBOQ}\APACrefatitle {Bayesian inversion for finite fault earthquake source
  models I—Theory and algorithm} {Bayesian inversion for finite fault
  earthquake source models i—theory and algorithm}.{\BBCQ}
\newblock
\APACjournalVolNumPages{Geophysical Journal International}{194}{3}{1701--1726}.
\PrintBackRefs{\CurrentBib}

\bibitem [\protect \citeauthoryear {%
Moreno%
, Rosenau%
\BCBL {}\ \BBA {} Oncken%
}{%
Moreno%
\ \protect \BOthers {.}}{%
{\protect \APACyear {2010}}%
}]{%
moreno20102010}
\APACinsertmetastar {%
moreno20102010}%
\begin{APACrefauthors}%
Moreno, M.%
, Rosenau, M.%
\BCBL {}\ \BBA {} Oncken, O.%
\end{APACrefauthors}%
\unskip\
\newblock
\APACrefYearMonthDay{2010}{}{}.
\newblock
{\BBOQ}\APACrefatitle {2010 Maule earthquake slip correlates with pre-seismic
  locking of Andean subduction zone} {2010 maule earthquake slip correlates
  with pre-seismic locking of andean subduction zone}.{\BBCQ}
\newblock
\APACjournalVolNumPages{Nature}{467}{7312}{198--202}.
\PrintBackRefs{\CurrentBib}

\bibitem [\protect \citeauthoryear {%
Nalbant%
, McCloskey%
, Steacy%
\BCBL {}\ \BBA {} Barka%
}{%
Nalbant%
\ \protect \BOthers {.}}{%
{\protect \APACyear {2002}}%
}]{%
nalbant2002stress}
\APACinsertmetastar {%
nalbant2002stress}%
\begin{APACrefauthors}%
Nalbant, S\BPBI S.%
, McCloskey, J.%
, Steacy, S.%
\BCBL {}\ \BBA {} Barka, A\BPBI A.%
\end{APACrefauthors}%
\unskip\
\newblock
\APACrefYearMonthDay{2002}{}{}.
\newblock
{\BBOQ}\APACrefatitle {Stress accumulation and increased seismic risk in
  eastern Turkey} {Stress accumulation and increased seismic risk in eastern
  turkey}.{\BBCQ}
\newblock
\APACjournalVolNumPages{Earth and Planetary Science
  Letters}{195}{3-4}{291--298}.
\PrintBackRefs{\CurrentBib}

\bibitem [\protect \citeauthoryear {%
Nocquet%
}{%
Nocquet%
}{%
{\protect \APACyear {2012}}%
}]{%
nocquet2012present}
\APACinsertmetastar {%
nocquet2012present}%
\begin{APACrefauthors}%
Nocquet, J\BHBI M.%
\end{APACrefauthors}%
\unskip\
\newblock
\APACrefYearMonthDay{2012}{}{}.
\newblock
{\BBOQ}\APACrefatitle {Present-day kinematics of the Mediterranean: A
  comprehensive overview of GPS results} {Present-day kinematics of the
  mediterranean: A comprehensive overview of gps results}.{\BBCQ}
\newblock
\APACjournalVolNumPages{Tectonophysics}{579}{}{220--242}.
\PrintBackRefs{\CurrentBib}

\bibitem [\protect \citeauthoryear {%
Nocquet%
}{%
Nocquet%
}{%
{\protect \APACyear {2018}}%
}]{%
nocquet2018stochastic}
\APACinsertmetastar {%
nocquet2018stochastic}%
\begin{APACrefauthors}%
Nocquet, J\BHBI M.%
\end{APACrefauthors}%
\unskip\
\newblock
\APACrefYearMonthDay{2018}{}{}.
\newblock
{\BBOQ}\APACrefatitle {Stochastic static fault slip inversion from geodetic
  data with non-negativity and bound constraints} {Stochastic static fault slip
  inversion from geodetic data with non-negativity and bound
  constraints}.{\BBCQ}
\newblock
\APACjournalVolNumPages{Geophysical Journal International}{214}{1}{366--385}.
\PrintBackRefs{\CurrentBib}

\bibitem [\protect \citeauthoryear {%
Nocquet%
, Calais%
, Altamimi%
, Sillard%
\BCBL {}\ \BBA {} Boucher%
}{%
Nocquet%
\ \protect \BOthers {.}}{%
{\protect \APACyear {2001}}%
}]{%
nocquet2001intraplate}
\APACinsertmetastar {%
nocquet2001intraplate}%
\begin{APACrefauthors}%
Nocquet, J\BHBI M.%
, Calais, E.%
, Altamimi, Z.%
, Sillard, P.%
\BCBL {}\ \BBA {} Boucher, C.%
\end{APACrefauthors}%
\unskip\
\newblock
\APACrefYearMonthDay{2001}{}{}.
\newblock
{\BBOQ}\APACrefatitle {Intraplate deformation in western Europe deduced from an
  analysis of the International Terrestrial Reference Frame 1997 (ITRF97)
  velocity field} {Intraplate deformation in western europe deduced from an
  analysis of the international terrestrial reference frame 1997 (itrf97)
  velocity field}.{\BBCQ}
\newblock
\APACjournalVolNumPages{Journal of Geophysical Research: Solid
  Earth}{106}{B6}{11239--11257}.
\PrintBackRefs{\CurrentBib}

\bibitem [\protect \citeauthoryear {%
Nocquet%
\ \protect \BOthers {.}}{%
Nocquet%
\ \protect \BOthers {.}}{%
{\protect \APACyear {2017}}%
}]{%
nocquet2017supercycle}
\APACinsertmetastar {%
nocquet2017supercycle}%
\begin{APACrefauthors}%
Nocquet, J\BHBI M.%
, Jarrin, P.%
, Vall{\'e}e, M.%
, Mothes, P.%
, Grandin, R.%
, Rolandone, F.%
\BDBL {}others%
\end{APACrefauthors}%
\unskip\
\newblock
\APACrefYearMonthDay{2017}{}{}.
\newblock
{\BBOQ}\APACrefatitle {Supercycle at the Ecuadorian subduction zone revealed
  after the 2016 Pedernales earthquake} {Supercycle at the ecuadorian
  subduction zone revealed after the 2016 pedernales earthquake}.{\BBCQ}
\newblock
\APACjournalVolNumPages{Nature Geoscience}{10}{2}{145--149}.
\PrintBackRefs{\CurrentBib}

\bibitem [\protect \citeauthoryear {%
Okada%
}{%
Okada%
}{%
{\protect \APACyear {1985}}%
}]{%
okada1985surface}
\APACinsertmetastar {%
okada1985surface}%
\begin{APACrefauthors}%
Okada, Y.%
\end{APACrefauthors}%
\unskip\
\newblock
\APACrefYearMonthDay{1985}{}{}.
\newblock
{\BBOQ}\APACrefatitle {Surface deformation due to shear and tensile faults in a
  half-space} {Surface deformation due to shear and tensile faults in a
  half-space}.{\BBCQ}
\newblock
\APACjournalVolNumPages{Bulletin of the seismological society of
  America}{75}{4}{1135--1154}.
\PrintBackRefs{\CurrentBib}

\bibitem [\protect \citeauthoryear {%
Ozener%
\ \protect \BOthers {.}}{%
Ozener%
\ \protect \BOthers {.}}{%
{\protect \APACyear {2010}}%
}]{%
ozener2010kinematics}
\APACinsertmetastar {%
ozener2010kinematics}%
\begin{APACrefauthors}%
Ozener, H.%
, Arpat, E.%
, Ergintav, S.%
, Dogru, A.%
, Cakmak, R.%
, Turgut, B.%
\BCBL {}\ \BBA {} Dogan, U.%
\end{APACrefauthors}%
\unskip\
\newblock
\APACrefYearMonthDay{2010}{}{}.
\newblock
{\BBOQ}\APACrefatitle {Kinematics of the eastern part of the North Anatolian
  Fault Zone} {Kinematics of the eastern part of the north anatolian fault
  zone}.{\BBCQ}
\newblock
\APACjournalVolNumPages{Journal of geodynamics}{49}{3-4}{141--150}.
\PrintBackRefs{\CurrentBib}

\bibitem [\protect \citeauthoryear {%
Pousse-Beltran%
\ \protect \BOthers {.}}{%
Pousse-Beltran%
\ \protect \BOthers {.}}{%
{\protect \APACyear {2020}}%
}]{%
pousse20202020}
\APACinsertmetastar {%
pousse20202020}%
\begin{APACrefauthors}%
Pousse-Beltran, L.%
, Nissen, E.%
, Bergman, E\BPBI A.%
, Cambaz, M\BPBI D.%
, Gaudreau, {\'E}.%
, Karas{\"o}zen, E.%
\BCBL {}\ \BBA {} Tan, F.%
\end{APACrefauthors}%
\unskip\
\newblock
\APACrefYearMonthDay{2020}{}{}.
\newblock
{\BBOQ}\APACrefatitle {The 2020 Mw 6.8 Elaz{\i}{\u{g}} (Turkey) earthquake
  reveals rupture behavior of the East Anatolian Fault} {The 2020 mw 6.8
  elaz{\i}{\u{g}} (turkey) earthquake reveals rupture behavior of the east
  anatolian fault}.{\BBCQ}
\newblock
\APACjournalVolNumPages{Geophysical Research Letters}{}{}{e2020GL088136}.
\PrintBackRefs{\CurrentBib}

\bibitem [\protect \citeauthoryear {%
Protti%
\ \protect \BOthers {.}}{%
Protti%
\ \protect \BOthers {.}}{%
{\protect \APACyear {2014}}%
}]{%
protti2014nicoya}
\APACinsertmetastar {%
protti2014nicoya}%
\begin{APACrefauthors}%
Protti, M.%
, Gonz{\'a}lez, V.%
, Newman, A\BPBI V.%
, Dixon, T\BPBI H.%
, Schwartz, S\BPBI Y.%
, Marshall, J\BPBI S.%
\BDBL {}Owen, S\BPBI E.%
\end{APACrefauthors}%
\unskip\
\newblock
\APACrefYearMonthDay{2014}{}{}.
\newblock
{\BBOQ}\APACrefatitle {Nicoya earthquake rupture anticipated by geodetic
  measurement of the locked plate interface} {Nicoya earthquake rupture
  anticipated by geodetic measurement of the locked plate interface}.{\BBCQ}
\newblock
\APACjournalVolNumPages{Nature Geoscience}{7}{2}{117--121}.
\PrintBackRefs{\CurrentBib}

\bibitem [\protect \citeauthoryear {%
Ragon%
\ \protect \BOthers {.}}{%
Ragon%
\ \protect \BOthers {.}}{%
{\protect \APACyear {2019b}}%
}]{%
ragon2019joint}
\APACinsertmetastar {%
ragon2019joint}%
\begin{APACrefauthors}%
Ragon, T.%
, Sladen, A.%
, Bletery, Q.%
, Vergnolle, M.%
, Cavali{\'e}, O.%
, Avallone, A.%
\BDBL {}Delouis, B.%
\end{APACrefauthors}%
\unskip\
\newblock
\APACrefYearMonthDay{2019b}{}{}.
\newblock
{\BBOQ}\APACrefatitle {Joint Inversion of Coseismic and Early Postseismic Slip
  to Optimize the Information Content in Geodetic Data: Application to the 2009
  M w 6.3 L'Aquila Earthquake, Central Italy} {Joint inversion of coseismic and
  early postseismic slip to optimize the information content in geodetic data:
  Application to the 2009 m w 6.3 l'aquila earthquake, central italy}.{\BBCQ}
\newblock
\APACjournalVolNumPages{Journal of Geophysical Research: Solid
  Earth}{124}{10}{10522--10543}.
\PrintBackRefs{\CurrentBib}

\bibitem [\protect \citeauthoryear {%
Ragon%
, Sladen%
\BCBL {}\ \BBA {} Simons%
}{%
Ragon%
\ \protect \BOthers {.}}{%
{\protect \APACyear {2018}}%
}]{%
ragon2018accounting}
\APACinsertmetastar {%
ragon2018accounting}%
\begin{APACrefauthors}%
Ragon, T.%
, Sladen, A.%
\BCBL {}\ \BBA {} Simons, M.%
\end{APACrefauthors}%
\unskip\
\newblock
\APACrefYearMonthDay{2018}{}{}.
\newblock
{\BBOQ}\APACrefatitle {Accounting for uncertain fault geometry in earthquake
  source inversions--I: theory and simplified application} {Accounting for
  uncertain fault geometry in earthquake source inversions--i: theory and
  simplified application}.{\BBCQ}
\newblock
\APACjournalVolNumPages{Geophysical Journal International}{214}{2}{1174--1190}.
\PrintBackRefs{\CurrentBib}

\bibitem [\protect \citeauthoryear {%
Ragon%
, Sladen%
\BCBL {}\ \BBA {} Simons%
}{%
Ragon%
\ \protect \BOthers {.}}{%
{\protect \APACyear {2019a}}%
}]{%
ragon2019accounting}
\APACinsertmetastar {%
ragon2019accounting}%
\begin{APACrefauthors}%
Ragon, T.%
, Sladen, A.%
\BCBL {}\ \BBA {} Simons, M.%
\end{APACrefauthors}%
\unskip\
\newblock
\APACrefYearMonthDay{2019a}{}{}.
\newblock
{\BBOQ}\APACrefatitle {Accounting for uncertain fault geometry in earthquake
  source inversions--II: application to the M w 6.2 Amatrice earthquake,
  central Italy} {Accounting for uncertain fault geometry in earthquake source
  inversions--ii: application to the m w 6.2 amatrice earthquake, central
  italy}.{\BBCQ}
\newblock
\APACjournalVolNumPages{Geophysical Journal International}{218}{1}{689--707}.
\PrintBackRefs{\CurrentBib}

\bibitem [\protect \citeauthoryear {%
Reilinger%
\ \BBA {} McClusky%
}{%
Reilinger%
\ \BBA {} McClusky%
}{%
{\protect \APACyear {2011}}%
}]{%
reilinger2011nubia}
\APACinsertmetastar {%
reilinger2011nubia}%
\begin{APACrefauthors}%
Reilinger, R.%
\BCBT {}\ \BBA {} McClusky, S.%
\end{APACrefauthors}%
\unskip\
\newblock
\APACrefYearMonthDay{2011}{}{}.
\newblock
{\BBOQ}\APACrefatitle {Nubia--Arabia--Eurasia plate motions and the dynamics of
  Mediterranean and Middle East tectonics} {Nubia--arabia--eurasia plate
  motions and the dynamics of mediterranean and middle east tectonics}.{\BBCQ}
\newblock
\APACjournalVolNumPages{Geophysical Journal International}{186}{3}{971--979}.
\PrintBackRefs{\CurrentBib}

\bibitem [\protect \citeauthoryear {%
Reilinger%
\ \protect \BOthers {.}}{%
Reilinger%
\ \protect \BOthers {.}}{%
{\protect \APACyear {2006}}%
}]{%
reilinger2006gps}
\APACinsertmetastar {%
reilinger2006gps}%
\begin{APACrefauthors}%
Reilinger, R.%
, McClusky, S.%
, Vernant, P.%
, Lawrence, S.%
, Ergintav, S.%
, Cakmak, R.%
\BDBL {}others%
\end{APACrefauthors}%
\unskip\
\newblock
\APACrefYearMonthDay{2006}{}{}.
\newblock
{\BBOQ}\APACrefatitle {GPS constraints on continental deformation in the
  Africa-Arabia-Eurasia continental collision zone and implications for the
  dynamics of plate interactions} {Gps constraints on continental deformation
  in the africa-arabia-eurasia continental collision zone and implications for
  the dynamics of plate interactions}.{\BBCQ}
\newblock
\APACjournalVolNumPages{Journal of Geophysical Research: Solid
  Earth}{111}{B5}{}.
\PrintBackRefs{\CurrentBib}

\bibitem [\protect \citeauthoryear {%
Stein%
, Barka%
\BCBL {}\ \BBA {} Dieterich%
}{%
Stein%
\ \protect \BOthers {.}}{%
{\protect \APACyear {1997}}%
}]{%
stein1997progressive}
\APACinsertmetastar {%
stein1997progressive}%
\begin{APACrefauthors}%
Stein, R\BPBI S.%
, Barka, A\BPBI A.%
\BCBL {}\ \BBA {} Dieterich, J\BPBI H.%
\end{APACrefauthors}%
\unskip\
\newblock
\APACrefYearMonthDay{1997}{}{}.
\newblock
{\BBOQ}\APACrefatitle {Progressive failure on the North Anatolian fault since
  1939 by earthquake stress triggering} {Progressive failure on the north
  anatolian fault since 1939 by earthquake stress triggering}.{\BBCQ}
\newblock
\APACjournalVolNumPages{Geophysical Journal International}{128}{3}{594--604}.
\PrintBackRefs{\CurrentBib}

\bibitem [\protect \citeauthoryear {%
Sudhaus%
\ \BBA {} Sigurj{\'o}n%
}{%
Sudhaus%
\ \BBA {} Sigurj{\'o}n%
}{%
{\protect \APACyear {2009}}%
}]{%
sudhaus2009improved}
\APACinsertmetastar {%
sudhaus2009improved}%
\begin{APACrefauthors}%
Sudhaus, H.%
\BCBT {}\ \BBA {} Sigurj{\'o}n, J.%
\end{APACrefauthors}%
\unskip\
\newblock
\APACrefYearMonthDay{2009}{}{}.
\newblock
{\BBOQ}\APACrefatitle {Improved source modelling through combined use of InSAR
  and GPS under consideration of correlated data errors: application to the
  June 2000 Kleifarvatn earthquake, Iceland} {Improved source modelling through
  combined use of insar and gps under consideration of correlated data errors:
  application to the june 2000 kleifarvatn earthquake, iceland}.{\BBCQ}
\newblock
\APACjournalVolNumPages{Geophysical Journal International}{176}{2}{389--404}.
\PrintBackRefs{\CurrentBib}

\bibitem [\protect \citeauthoryear {%
Tan%
, Tapirdamaz%
\BCBL {}\ \BBA {} Y{\"o}r{\"u}k%
}{%
Tan%
\ \protect \BOthers {.}}{%
{\protect \APACyear {2008}}%
}]{%
tan2008earthquake}
\APACinsertmetastar {%
tan2008earthquake}%
\begin{APACrefauthors}%
Tan, O.%
, Tapirdamaz, M\BPBI C.%
\BCBL {}\ \BBA {} Y{\"o}r{\"u}k, A.%
\end{APACrefauthors}%
\unskip\
\newblock
\APACrefYearMonthDay{2008}{}{}.
\newblock
{\BBOQ}\APACrefatitle {The earthquake catalogues for Turkey} {The earthquake
  catalogues for turkey}.{\BBCQ}
\newblock
\APACjournalVolNumPages{Turkish Journal of Earth Sciences}{17}{2}{405--418}.
\PrintBackRefs{\CurrentBib}

\bibitem [\protect \citeauthoryear {%
Tarantola%
\ \BBA {} Valette%
}{%
Tarantola%
\ \BBA {} Valette%
}{%
{\protect \APACyear {1982}}%
}]{%
tarantola1982inverse}
\APACinsertmetastar {%
tarantola1982inverse}%
\begin{APACrefauthors}%
Tarantola, A.%
\BCBT {}\ \BBA {} Valette, B.%
\end{APACrefauthors}%
\unskip\
\newblock
\APACrefYearMonthDay{1982}{}{}.
\newblock
{\BBOQ}\APACrefatitle {Inverse problems= quest for information} {Inverse
  problems= quest for information}.{\BBCQ}
\newblock
\APACjournalVolNumPages{Journal of geophysics}{50}{1}{159--170}.
\PrintBackRefs{\CurrentBib}

\bibitem [\protect \citeauthoryear {%
Tatar%
\ \protect \BOthers {.}}{%
Tatar%
\ \protect \BOthers {.}}{%
{\protect \APACyear {2012}}%
}]{%
tatar2012crustal}
\APACinsertmetastar {%
tatar2012crustal}%
\begin{APACrefauthors}%
Tatar, O.%
, Poyraz, F.%
, G{\"u}rsoy, H.%
, Cakir, Z.%
, Ergintav, S.%
, Akp{\i}nar, Z.%
\BDBL {}others%
\end{APACrefauthors}%
\unskip\
\newblock
\APACrefYearMonthDay{2012}{}{}.
\newblock
{\BBOQ}\APACrefatitle {Crustal deformation and kinematics of the Eastern Part
  of the North Anatolian Fault Zone (Turkey) from GPS measurements} {Crustal
  deformation and kinematics of the eastern part of the north anatolian fault
  zone (turkey) from gps measurements}.{\BBCQ}
\newblock
\APACjournalVolNumPages{Tectonophysics}{518}{}{55--62}.
\PrintBackRefs{\CurrentBib}

\bibitem [\protect \citeauthoryear {%
T{\"u}rko{\u{g}}lu%
, Unsworth%
, Bulut%
\BCBL {}\ \BBA {} {\c{C}}a{\u{g}}lar%
}{%
T{\"u}rko{\u{g}}lu%
\ \protect \BOthers {.}}{%
{\protect \APACyear {2015}}%
}]{%
turkouglu2015crustal}
\APACinsertmetastar {%
turkouglu2015crustal}%
\begin{APACrefauthors}%
T{\"u}rko{\u{g}}lu, E.%
, Unsworth, M.%
, Bulut, F.%
\BCBL {}\ \BBA {} {\c{C}}a{\u{g}}lar, {\.I}.%
\end{APACrefauthors}%
\unskip\
\newblock
\APACrefYearMonthDay{2015}{}{}.
\newblock
{\BBOQ}\APACrefatitle {Crustal structure of the North Anatolian and East
  Anatolian Fault Systems from magnetotelluric data} {Crustal structure of the
  north anatolian and east anatolian fault systems from magnetotelluric
  data}.{\BBCQ}
\newblock
\APACjournalVolNumPages{Physics of the Earth and Planetary
  Interiors}{241}{}{1--14}.
\PrintBackRefs{\CurrentBib}

\bibitem [\protect \citeauthoryear {%
USGS%
}{%
USGS%
}{%
{\protect \APACyear {2020}}%
}]{%
usgs}
\APACinsertmetastar {%
usgs}%
\begin{APACrefauthors}%
USGS.%
\end{APACrefauthors}%
\unskip\
\newblock
\APACrefYearMonthDay{2020}{}{}.
\newblock
{\BBOQ}\APACrefatitle {Finite fault model for the 01 24 2020 M 6.7 earthquake,
  Turkey} {Finite fault model for the 01 24 2020 m 6.7 earthquake,
  turkey}.{\BBCQ}
\newblock
\APACjournalVolNumPages{USGS finite-fault solution}{}{}{}.
\newblock
\begin{APACrefURL}
  \url{http://earthquake.usgs.gov/earthquakes/eventpage/us60007ewc/finite-fault?source=us&code=us60007ewc}
  \end{APACrefURL}
\PrintBackRefs{\CurrentBib}

\bibitem [\protect \citeauthoryear {%
Van~Wijk%
, Axen%
\BCBL {}\ \BBA {} Abera%
}{%
Van~Wijk%
\ \protect \BOthers {.}}{%
{\protect \APACyear {2017}}%
}]{%
van2017initiation}
\APACinsertmetastar {%
van2017initiation}%
\begin{APACrefauthors}%
Van~Wijk, J.%
, Axen, G.%
\BCBL {}\ \BBA {} Abera, R.%
\end{APACrefauthors}%
\unskip\
\newblock
\APACrefYearMonthDay{2017}{}{}.
\newblock
{\BBOQ}\APACrefatitle {Initiation, evolution and extinction of pull-apart
  basins: Implications for opening of the Gulf of California} {Initiation,
  evolution and extinction of pull-apart basins: Implications for opening of
  the gulf of california}.{\BBCQ}
\newblock
\APACjournalVolNumPages{Tectonophysics}{719}{}{37--50}.
\PrintBackRefs{\CurrentBib}

\bibitem [\protect \citeauthoryear {%
Wang%
, Liu%
, Ye%
, Cao%
\BCBL {}\ \BBA {} Jing%
}{%
Wang%
\ \protect \BOthers {.}}{%
{\protect \APACyear {2017}}%
}]{%
wang2017strain}
\APACinsertmetastar {%
wang2017strain}%
\begin{APACrefauthors}%
Wang, H.%
, Liu, M.%
, Ye, J.%
, Cao, J.%
\BCBL {}\ \BBA {} Jing, Y.%
\end{APACrefauthors}%
\unskip\
\newblock
\APACrefYearMonthDay{2017}{}{}.
\newblock
{\BBOQ}\APACrefatitle {Strain partitioning and stress perturbation around
  stepovers and bends of strike-slip faults: Numerical results} {Strain
  partitioning and stress perturbation around stepovers and bends of
  strike-slip faults: Numerical results}.{\BBCQ}
\newblock
\APACjournalVolNumPages{Tectonophysics}{721}{}{211--226}.
\PrintBackRefs{\CurrentBib}

\bibitem [\protect \citeauthoryear {%
Wesnousky%
}{%
Wesnousky%
}{%
{\protect \APACyear {2006}}%
}]{%
wesnousky2006predicting}
\APACinsertmetastar {%
wesnousky2006predicting}%
\begin{APACrefauthors}%
Wesnousky, S\BPBI G.%
\end{APACrefauthors}%
\unskip\
\newblock
\APACrefYearMonthDay{2006}{}{}.
\newblock
{\BBOQ}\APACrefatitle {Predicting the endpoints of earthquake ruptures}
  {Predicting the endpoints of earthquake ruptures}.{\BBCQ}
\newblock
\APACjournalVolNumPages{Nature}{444}{7117}{358--360}.
\PrintBackRefs{\CurrentBib}

\end{thebibliography}

%
%
%
%
%

\end{document}